\documentclass[10pt,twocolumn,letterpaper]{article}

\usepackage{cvpr}              

\definecolor{cvprblue}{rgb}{0.21,0.49,0.74}
\usepackage[pagebackref,breaklinks,colorlinks,allcolors=cvprblue]{hyperref}

\usepackage{booktabs}

\def\paperID{14590} 
\def\confName{CVPR}
\def\confYear{2025}

\title{From Head to Tail: Efficient Black-box Model Inversion Attack\\via Long-tailed Learning}

\author{
Ziang Li\textsuperscript{1,2},
Hongguang Zhang\textsuperscript{1},
Juan Wang\textsuperscript{1}\thanks{Corresponding author.},
Meihui Chen\textsuperscript{3},
Hongxin Hu\textsuperscript{4}, \\
Wenzhe Yi\textsuperscript{1},
Xiaoyang Xu\textsuperscript{1},
Mengda Yang\textsuperscript{1},
Chenjun Ma\textsuperscript{1}\\
\textsuperscript{1} Key Laboratory of Aerospace Information Security and Trusted Computing, Ministry of Education, \\ School of Cyber Science and Engineering, Wuhan University\\
\textsuperscript{2} Shanghai Innovation Institute ~~
\textsuperscript{3} Ant Group\\
\textsuperscript{4} Department of Computer Science and Engineering, University at Buffalo, SUNY\\
{\tt\small {\textsuperscript{1}\{ziangli, hongguangz, jwang, wenzhey, xiaoyangx, mengday, mcj123\}@whu.edu.cn}}\\
{\tt\small {\textsuperscript{3}chenmeihui.cmh@antgroup.com}}~~{\tt\small {\textsuperscript{4}hongxinh@buffalo.edu}}
}

\begin{document}
\maketitle

\begin{abstract}
Model Inversion Attacks (MIAs) aim to reconstruct private training data from models, leading to privacy leakage, particularly in facial recognition systems. Although many studies have enhanced the effectiveness of white-box MIAs, less attention has been paid to improving efficiency and utility under limited attacker capabilities. Existing black-box MIAs necessitate an impractical number of queries, incurring significant overhead. Therefore, we analyze the limitations of existing MIAs and introduce \textbf{S}urrogate \textbf{M}odel-based \textbf{I}nversion with \textbf{L}ong-tailed \textbf{E}nhancement (\textbf{SMILE}), a high-resolution oriented and query-efficient MIA for the black-box setting. We begin by analyzing the initialization of MIAs from a data distribution perspective and propose a long-tailed surrogate training method to obtain high-quality initial points. We then enhance the attack's effectiveness by employing the gradient-free black-box optimization algorithm selected by NGOpt. Our experiments show that \textbf{SMILE} outperforms existing state-of-the-art black-box MIAs while requiring only about 5\% of the query overhead. Our code is available at \url{https://github.com/L1ziang/SMILE}.
\vspace{-20pt}

\end{abstract}    
\section{Introduction}
\vspace{-3pt}
\label{sec:intro}
In recent years, deep neural networks have been widely applied in information-sensitive fields such as healthcare~\cite{richens2020improving}, finance~\cite{ozbayoglu2020deep} and image editing~\cite{wang2024high,wang2024magicface}, raising increasing concerns regarding privacy~\cite{shokri2017membership,geiping2020inverting,li2023gan,yang2022measuring,xu2024stealthy}, especially training data privacy. A significant privacy threat is the model inversion attack (MIA), which aims to expose information about the training data of models. Typically, MIAs reconstruct private instances from face recognition models.

\vspace{-10pt}
\paragraph{Model Inversion Attack.}~\cite{fredrikson2014privacy} first proposed MIA, targeting linear regression models. Recent works have primarily focused on visual face recognition models, performing MIAs in the white-box setting, where the attacker has access to the target model parameters. GMI~\cite{zhang2020secret} first utilized the prior knowledge provided by GANs~\cite{goodfellow2014generative}, formalizing MIA as an optimization problem. Based on this paradigm, subsequent studies have further improved the effect of reconstruction~\cite{chen2021knowledge,wang2021variational,yuan2023pseudo,nguyen2023re,struppek2022plug,an2022mirror}. Specifically, Mirror~\cite{an2022mirror} utilizes StyleGAN~\cite{karras2019style} to achieve high-resolution reconstruction results on open-source pre-trained models. In black-box MIA, attackers can obtain the model output. RLBMI~\cite{han2023reinforcement} combines reinforcement learning, and Mirror~\cite{an2022mirror} employs genetic algorithms to solve this optimization problem. \cite{kahla2022label,nguyen2024label} focus on the more challenging setting where attackers can only access hard labels.

\vspace{-12pt}
\paragraph{Long-tailed Learning.} The long-tail distribution is typically reflected in the fact that a few individuals contribute the most, dominating the dataset as head classes, while most classes, called tail classes, have few data samples~\cite{yang2022survey,yue2024revisiting}. Models trained on long-tail distributions often exhibit bias towards head classes, leading to suboptimal performance on tail classes. Common mitigation solutions such as resampling~\cite{drummond2003c4,chawla2002smote}, reweighting~\cite{huang2016learning,cui2019class}, margin modifications~\cite{cao2019learning,menon2020long}, and ensemble learning~\cite{cai2021ace,wang2020long} are employed to improve long-tailed recognition performance.

In this paper, we propose an efficient black-box MIA, named \textbf{S}urrogate \textbf{M}odel-based \textbf{I}nversion with \textbf{L}ong-tailed \textbf{E}nhancement (\textit{SMILE}). We analyze the limitations of existing black-box MIAs from two perspectives: the targeted datasets and the number of queries. To address these limitations, we follow the principle that \textit{high-quality initial points are crucial for optimization problems} and conduct detailed data analysis and visualization of the logits output by the target model. By leveraging the proposed \textbf{Long-tailed surrogate training} method, we construct high-quality local surrogate models with an extremely limited number of queries, which significantly reduces the complexity of the subsequent black-box optimization process. In summary, our contributions are as follows:
\begin{itemize}
    \item We perform a fine-grained data analysis on the output logits from open-source pre-trained models. We find that the number of samples across classes exhibits an extreme long-tail distribution and discuss possible mitigation strategies.
    \item We model the training of surrogate models as a long-tailed recognition problem. By utilizing the proposed long-tailed surrogate training method, we obtain high-quality surrogate models with extremely limited queries. This provides advantageous initial points for subsequent optimization processes.
    \item We propose \textit{SMILE},  an efficient black-box MIA, which integrates long-tailed surrogate training and the black-box optimization algorithm selected by NGOpt. Experiments demonstrate that \textit{SMILE} surpasses existing state-of-the-art black-box MIAs, while requiring only about 5\% of the query overhead.
\end{itemize}

\section{Motivation}
\label{sec:motivation}
Although MIAs have been extensively studied, state-of-the-art black-box MIAs still face significant limitations that seriously affect their practicability and effectiveness. 

\subsection{The targeted datasets}

One primary limitation stems from the characteristics of the targeted private training datasets. Similarly to most MIAs, including GMI~\cite{zhang2020secret}, KEDMI~\cite{chen2021knowledge}, VMI~\cite{wang2021variational}, PLGMI~\cite{yuan2023pseudo}, LOMMA~\cite{nguyen2023re}, BREPMI~\cite{kahla2022label}, and LOKT~\cite{nguyen2024label}, the datasets targeted by RLBMI~\cite{han2023reinforcement} feature low-resolution images and a limited number of IDs. These datasets generally undergo alignment, clipping, and resizing to a uniform size of $64 \times 64$. Examples include CelebA~\cite{liu2015deep}, which contains 1,000 private IDs out of 10,177 total IDs; FaceScrub~\cite{ng2014data}, which contains 200 private IDs out of 530 total IDs; and PubFig83~\cite{pinto2011scaling}, with 50 private IDs out of 83 total IDs. Moreover, in the main experiments, GAN models trained on samples from the remaining IDs serve as image priors. This typical setting reduces the authenticity and complexity of the MIA, because it simplifies the process of MIAs via GAN latent space optimization to disclose private facial features. 
Therefore, we follow Mirror's setting~\cite{an2022mirror} where: 


\begin{itemize}
    \item The target models are pre-trained models available online.
    \item The number of IDs in the private training data is massive.
    \item The synthetic data from pre-trained GAN models and the private training data originate from different distributions.
\end{itemize}
These alleviate the above limitation, as detailed in \cref{table:motivation_table_1}.

\subsection{The number of queries}

Another limitation arises from the number of queries to the black-box target model. In black-box attacks, query cost is a crucial metric, second only to the attack's success rate. One significant drawback of current black-box MIAs is the monetary cost caused by the large number of queries. For instance, utilizing a commercial face recognition API costs \$0.001 or more per call~\cite{Amazon,Microsoft,faceplusplus}. If an attacker requires over $100K$ queries to compromise the facial privacy of a target ID, the expense would amount to \$100. Furthermore, during an attack on a fixed ID, the optimization process of MIAs tends to generate numerous intermediate results with similar facial features. It is easy for the cloud service provider hosting the API to detect such large-scale and similar access requests and then interrupt the black-box MIAs. 

Regrettably, current black-box MIAs rely on executing a large number of queries. A general MIA process consists of latent vector sampling for initialization {\color[HTML]{9A0000}{\raisebox{-0.2ex}{\large \ding{172}}}}, iterative optimization for feature search {\color[HTML]{00008B}{\raisebox{-0.2ex}{\large \ding{173}}}}, and an optional post-processing step {\color[HTML]{006400}{\raisebox{-0.2ex}{\large \ding{174}}}}~\cite{qiu2024mibench}. For a selected target model, Step {\color[HTML]{9A0000}{\raisebox{-0.2ex}{\large \ding{172}}}} is required only once, while Step {\color[HTML]{00008B}{\raisebox{-0.2ex}{\large \ding{173}}}} and {\color[HTML]{006400}{\raisebox{-0.2ex}{\large \ding{174}}}} repeat for each targeted ID. As illustrated in \cref{table:motivation_table_2}, for Mirror-b~\cite{an2022mirror}, $100K$ samples are required for initialization. Each attack subsequently incurs an additional query overhead of $20K$. For RLBMI~\cite{han2023reinforcement}, although Step {\color[HTML]{9A0000}{\raisebox{-0.2ex}{\large \ding{172}}}} is not needed, it involves a query overhead of $80K$ in Step {\color[HTML]{00008B}{\raisebox{-0.2ex}{\large \ding{173}}}}. In \cref{table:motivation_table_2}, we also detail the query costs of white-box MIAs. Although the number of queries is less critical for white-box MIAs where attackers can obtain model parameters and run locally, a high number of initialization and optimization still increases the time and resource costs.
\begin{table}[h!]
  \centering
\resizebox{.47\textwidth}{!}{
\setlength{\tabcolsep}{2pt}
\begin{tabular}{cccccccc}
\toprule
MIAs     & Type & Step {\color[HTML]{9A0000}{\raisebox{-0.2ex}{\large \ding{172}}}} & Step {\color[HTML]{00008B}{\raisebox{-0.2ex}{\large \ding{173}}}} & Step {\color[HTML]{006400}{\raisebox{-0.2ex}{\large \ding{174}}}} & $N=1$ & $N=10$ & $N=50$ \\ \hline
Mirror-w~\cite{an2022mirror} & White-box    & 100K   & 20K    & —      & 120K      & 300K       & 1.1M       \\
PPA~\cite{struppek2022plug}      & White-box    & 5K     & 14K    & 5K     & 24K       & 195K       & 0.995M     \\ \hline
RLBMI~\cite{han2023reinforcement}    & Black-box    & 0K     & 80K    & —      & 80K       & 800K       & 4M         \\ 
Mirror-b~\cite{an2022mirror} & Black-box    & 100K   & 20K    & —      & 120K      & 300K       & 1.1M       \\
SMILE (ours)     & Black-box    & 2.5K   & 1K     & —      & 3.5K      & 12.5K      & 52.5K     \\ 
\bottomrule
\end{tabular}
}
\vspace{-6pt} 
  \caption{\footnotesize{\textbf{Statistics of the number of queries required by MIAs.} $N$ is the number of targeted IDs. The query overhead of our method is only about 5\% compared to the overhead of SOTA black-box MIAs.}}
\vspace{-15pt} 
  \label{table:motivation_table_2}
\end{table}

\begin{table*}[t!]
\centering
\resizebox{\textwidth}{!}{
\begin{tabular}{cccccccc}
\toprule
MIAs                                                                   & \multicolumn{6}{c}{Dateset ($N_{priv}$) \& Model Architecture \& Resolution}                                         & Image Priors                                                                \\ \hline
\multirow{3}{*}{\begin{tabular}[c]{@{}c@{}}~\cite{chen2021knowledge,han2023reinforcement,kahla2022label}\\ ~\cite{nguyen2024label,nguyen2023re,wang2021variational}\\ ~\cite{yuan2023pseudo,zhang2020secret}\end{tabular}}                                                 & \multicolumn{2}{c}{PubFig83 $({\color[HTML]{9A0000}50})$~\cite{pinto2011scaling}}     & \multicolumn{2}{c}{FaceScrub $({\color[HTML]{9A0000}200})$~\cite{ng2014data}} & \multicolumn{2}{c}{CelebA $({\color[HTML]{9A0000}1000})$~\cite{liu2015deep}} & \multirow{3}{*}{\begin{tabular}[c]{@{}c@{}}{\color[HTML]{9A0000}IID Data}\\ FFHQ~\cite{karras2019style}\end{tabular}} \\ \cline{2-7}
                                                                       & \multicolumn{6}{c}{VGG16~\cite{simonyan2014very}, ResNet152~\cite{he2016deep}, Facenet64~\cite{cheng2017know}}                                              &                                                                             \\
                                                                       & \multicolumn{6}{c}{$64 \times 64$}                                                                                       &                                                                             \\ \hline
\multirow{3}{*}{PPA~\cite{struppek2022plug}}                                                   & \multicolumn{2}{c}{\multirow{3}{*}{}} & \multicolumn{2}{c}{FaceScrub $({\color[HTML]{9A0000}530})$~\cite{ng2014data}} & \multicolumn{2}{c}{CelebA $({\color[HTML]{9A0000}1000})$~\cite{liu2015deep}} & \multirow{3}{*}{\begin{tabular}[c]{@{}c@{}}FFHQ~\cite{karras2019style}\\ MetFaces~\cite{karras2020training}\end{tabular}}     \\ \cline{4-7}
                                                                       & \multicolumn{2}{c}{}                  & \multicolumn{4}{c}{ResNet~\cite{he2016deep}, ResNeSt~\cite{zhang2022resnest}, DenseNet~\cite{huang2017densely} Series}                                               &                                                                             \\
                                                                       & \multicolumn{2}{c}{}                  & \multicolumn{4}{c}{$224 \times 224$}                                             &                                                                             \\ \hline
\multirow{3}{*}{\begin{tabular}[c]{@{}c@{}}Mirror~\cite{an2022mirror}\\ Ours\end{tabular}} & \multicolumn{2}{c}{VGGFace $({\color[HTML]{9A0000}2622})$~\cite{parkhi2015deep}}    & \multicolumn{2}{c}{VGGFace2 $({\color[HTML]{9A0000}9131})$~\cite{cao2018vggface2}} & \multicolumn{2}{c}{CASIA $({\color[HTML]{9A0000}10575})$~\cite{yi2014learning}} & \multirow{3}{*}{\begin{tabular}[c]{@{}c@{}}CelebA~\cite{liu2015deep}\\ FFHQ~\cite{karras2019style}\end{tabular}}      \\ \cline{2-7}
                                                                       & VGG16~\cite{VGG16}            & VGG16BN~\cite{VGG16}            & ResNet50~\cite{VGG16}        & InceptionV1~\cite{INCEPTIONv1}       & InceptionV1~\cite{INCEPTIONv1}      & SphereFace~\cite{SPHEREFACE}     &                                                                             \\
                                                                       & \multicolumn{2}{c}{$224 \times 224$}           & $224 \times 224$         & $160 \times 160$           & $160 \times 160$          & $112 \times 196$         &                                                                             \\ \bottomrule
\end{tabular}
}
\vspace{-4pt} 
\caption{\footnotesize{\textbf{The datasets and models involved in MIAs.} The number of privacy IDs highlighted in {\color[HTML]{9A0000}red}. Most MIAs focus on low-resolution scenes and utilizing independently and identically distributed (IID) data as image priors, with only partial experiments involving shifted priors from FFHQ. We target a large number of private IDs, using image priors from different distributions, which implies more difficult MIA.}}
\vspace{-12pt} 
\label{table:motivation_table_1}
\end{table*}

\subsection{Our solution}
From the perspective of minimizing query overhead, one might easily conclude that an efficient black-box MIA should incorporate: 
\begin{itemize}
  \setlength{\itemindent}{2em}
  \item[\textbf{1)}] An initialization conducive to the optimization.
  \item[\textbf{2)}] An efficient black-box optimization algorithm.
  \item[\textbf{3)}] A one-shot attack mechanism for a specific target ID.
\end{itemize}

Specifically, for Step {\color[HTML]{9A0000}{\raisebox{-0.2ex}{\large \ding{172}}}}, more private information regarding the target ID should be mined. For Step {\color[HTML]{00008B}{\raisebox{-0.2ex}{\large \ding{173}}}}, a black-box optimization algorithm suitable for searching the GAN latent space is required, preferably without Step {\color[HTML]{006400}{\raisebox{-0.2ex}{\large \ding{174}}}}. To this end, we first feed GAN-synthesized facial samples into pre-trained facial recognition models and perform data analysis on the logits. From this, we introduce a surrogate training method based on long-tailed learning. By attacking the local surrogate models, Step {\color[HTML]{9A0000}{\raisebox{-0.2ex}{\large \ding{172}}}} of our method provides high-quality initialization points for Step {\color[HTML]{00008B}{\raisebox{-0.2ex}{\large \ding{173}}}} under a strict query budget. Then, in Step {\color[HTML]{00008B}{\raisebox{-0.2ex}{\large \ding{173}}}}, we apply the black-box optimization algorithm selector NGOpt~\cite{liu2020versatile,meunier2021black}, provided by Nevergrad~\cite{rapin2018nevergrad}. NGOpt automatically selects a suitable gradient-free black-box optimization as the solver based on high-level problem information. Based on the above, our solution achieves superior black-box MIA effects under an extremely constrained query budget, as detailed in \cref{table:motivation_table_2}.

\vspace{-4pt}
\section{Methodology}
\label{sec:methodology}
\vspace{-1pt}


\subsection{Threat model}

MIAs aim to uncover private training data based on the target model \(M_t\), which is typically modeled as an optimization problem with auxiliary priors. 
The auxiliary priors, derived from the public dataset \(\mathcal{D}_{pub}\), share the same or similar data distribution with the private dataset \(\mathcal{D}_{priv}\).
To narrow the search space and facilitate the optimization process, the prior knowledge provided by \(\mathcal{D}_{pub}\) is compressed into a generative model \(G\), typically an attacker-self-trained or pre-trained GAN model. 

In the workflow of a GAN-based MIA, the attacker initially samples a set of latent vectors \(\mathcal{Z}_{init}=\{z_1, \ldots, z_{N_{Step-1}}\}\) from the standard Gaussian distribution and employs the GAN model to generate an initial sample pool \(\mathcal{X}_{init}=G(\mathcal{Z}_{init})=\{x_1, \ldots, x_{N_{Step-1}}\}\), \(N_{Step-1}\) is the number of samples in Step {\color[HTML]{9A0000}{\raisebox{-0.2ex}{\large \ding{172}}}}. These samples are then fed into the \(M_t\). Based on the output logits \(\mathcal{Y}_{init}=M_t(\mathcal{X}_{init})=\{y_1, \ldots, y_{N_{Step-1}}\}\), the attacker selects the candidate \(x\sim G(z)\) that optimally corresponds to the target ID \(c \in \mathcal{C}_{target}=\{1, \ldots, N_{target}\}\) — typically the one closer to the target individual in perceptual distance. The candidate \(z\) serves as an initialization and proceeds to the subsequent optimization process, formalized as:
\begin{align}
    \min_{z\in\mathcal{Z}_{init}}\ \mathcal{L}_{id}(M_t(G(z)), c)+\lambda\  \mathcal{R}(G,z).
\label{eq:step2}
\end{align}
\(\mathcal{R}\) is regularization and the optimization process iterates \(N_{Step-2}\) times to minimize the loss, and finally \(z^*\) is obtained. The attacker takes \(x^*=G(z^*)\) as the MIA result on \(c\). In the case of black-box MIA~\cite{an2022mirror,han2023reinforcement}, the attacker can query \(M_t\) but cannot access its parameters, so that ~\cref{eq:step2} cannot be optimized via backpropagation, which is the primary reason for the extensive number of queries required in black-box MIAs. We assume attackers know the application domain of the target model, but are not aware of its training method or model architecture. For practicality, in our main experiments, both the black-box target models and the GAN models are from open-source, pre-trained models. Additionally, in our setting, the attacker is able to obtain a model pre-trained on an arbitrary face dataset. This is feasible, as there are numerous such pre-trained models available on the Internet. We also further evaluate models trained for MIA defense purposes.

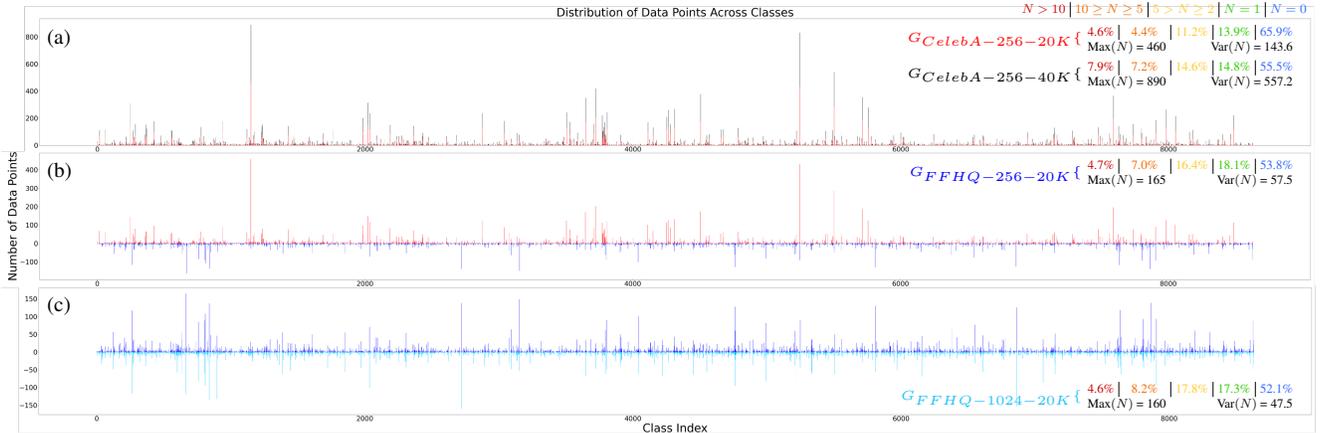
\begin{figure*}[h!]
\centering
\begin{overpic}[width=\textwidth]{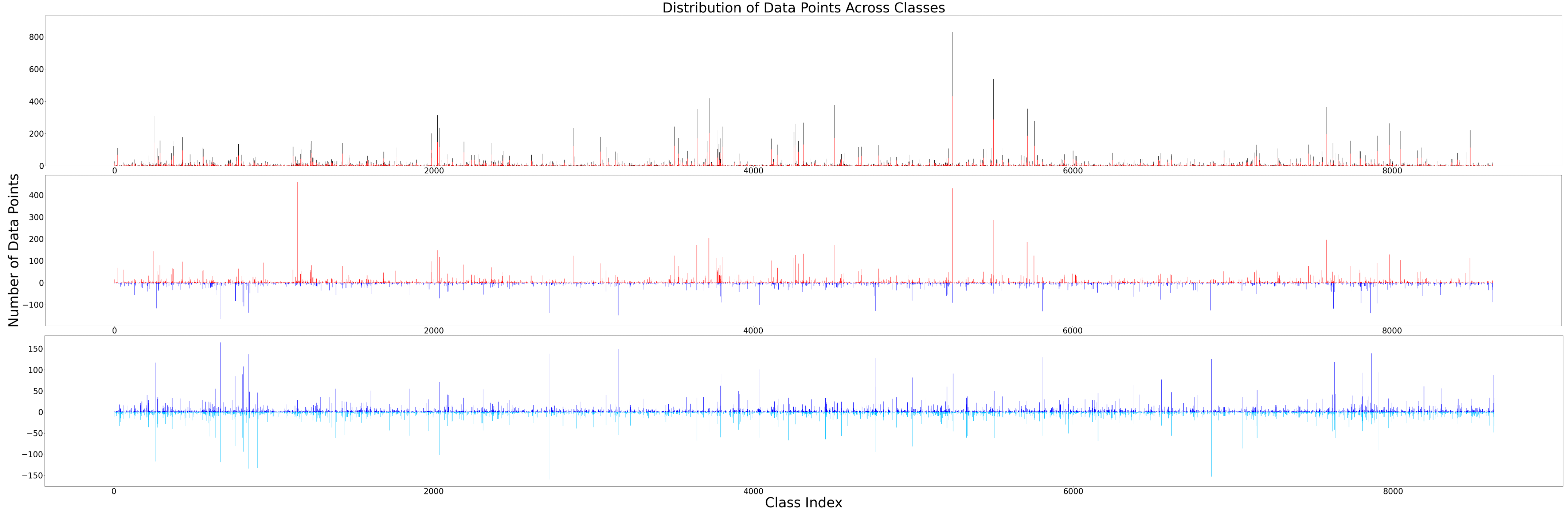} 

\put(76,32){ 
\scalebox{0.6}{
\setlength{\tabcolsep}{3pt}
\footnotesize{
\begin{tabular}{l|l|l|l|l}
{\color[HTML]{CB0000} \(N>10\)} & {\color[HTML]{F56B00} \(10\geq N \geq 5\)} & {\color[HTML]{FFCA2C} \(5>N \geq 2\)} & {\color[HTML]{32CB00} \(N=1\)} & {\color[HTML]{3166FF} \(N=0\)}
\end{tabular}
}}}

\put(81,29.7){ 
\scalebox{0.6}{
\setlength{\tabcolsep}{3pt}
\footnotesize{
\begin{tabular}{ccccc}
\multicolumn{1}{c|}{{\color[HTML]{CB0000} 4.6\%}} & \multicolumn{1}{c|}{{\color[HTML]{F56B00} 4.4\%}} & \multicolumn{1}{c|}{{\color[HTML]{FFCA2C} 11.2\%}} & \multicolumn{1}{c|}{{\color[HTML]{32CB00} 13.9\%}} & {\color[HTML]{3166FF} 65.9\%} \\
\multicolumn{2}{c}{\(\text{Max}(N)\) = 460}                                                                                                                           & \multicolumn{3}{r}{\(\text{Var}(N)\) = 143.6}                                                 
\end{tabular}
}}}

\put(81,27){ 
\scalebox{0.6}{
\setlength{\tabcolsep}{3pt}
\footnotesize{
\begin{tabular}{ccccc}
\multicolumn{1}{c|}{{\color[HTML]{CB0000} 7.9\%}} & \multicolumn{1}{c|}{{\color[HTML]{F56B00} 7.2\%}} & \multicolumn{1}{c|}{{\color[HTML]{FFCA2C} 14.6\%}} & \multicolumn{1}{c|}{{\color[HTML]{32CB00} 14.8\%}} & {\color[HTML]{3166FF} 55.5\%} \\
\multicolumn{2}{c}{\(\text{Max}(N)\) = 890}                                                                                                                           & \multicolumn{3}{r}{\(\text{Var}(N)\) = 557.2}                                                 
\end{tabular}
}}}

\put(81,19.5){ 
\scalebox{0.6}{
\setlength{\tabcolsep}{3pt}
\footnotesize{
\begin{tabular}{ccccc}
\multicolumn{1}{c|}{{\color[HTML]{CB0000} 4.7\%}} & \multicolumn{1}{c|}{{\color[HTML]{F56B00} 7.0\%}} & \multicolumn{1}{c|}{{\color[HTML]{FFCA2C} 16.4\%}} & \multicolumn{1}{c|}{{\color[HTML]{32CB00} 18.1\%}} & {\color[HTML]{3166FF} 53.8\%} \\
\multicolumn{2}{c}{\(\text{Max}(N)\) = 165}                                                                                                                           & \multicolumn{3}{r}{\(\text{Var}(N)\) = 57.5}                                                 
\end{tabular}
}}}

\put(81,2.5){ 
\scalebox{0.6}{
\setlength{\tabcolsep}{3pt}
\footnotesize{
\begin{tabular}{ccccc}
\multicolumn{1}{c|}{{\color[HTML]{CB0000} 4.6\%}} & \multicolumn{1}{c|}{{\color[HTML]{F56B00} 8.2\%}} & \multicolumn{1}{c|}{{\color[HTML]{FFCA2C} 17.8\%}} & \multicolumn{1}{c|}{{\color[HTML]{32CB00} 17.3\%}} & {\color[HTML]{3166FF} 52.1\%} \\
\multicolumn{2}{c}{\(\text{Max}(N)\) = 160}                                                                                                                           & \multicolumn{3}{r}{\(\text{Var}(N)\) = 47.5}                                                 
\end{tabular}
}}}

\put (3.5,29.65) {\footnotesize{(a)}}
\put (3.5,19.47) {\footnotesize{(b)}}
\put (3.5,9.3) {\footnotesize{(c)}}

\put (69,29.84) {\textcolor{red}{\tiny{\(G_{CelebA-256-20K}\)}\{ }}
\put (69,27.1) {\textcolor{black}{\tiny{\(G_{CelebA-256-40K}\)}\{ }}
\put (69.15,19.55) {\textcolor{blue}{\tiny{\(G_{FFHQ-256-20K}\)}\{ }}
\definecolor{customcolor}{RGB}{31,192,251}
\put (68.5,2.5) {\textcolor{customcolor}{\tiny{\(G_{FFHQ-1024-20K}\)}\{ }}

\end{overpic}
\vspace{-15pt} 
\caption{\footnotesize{\textbf{Visualization of sample distribution across classes.} $N$ denotes the number of samples in a single label. The colors of the bar chart indicate the GAN models used for sampling. From \textbf{(a)}, it can be seen that with the same image prior, doubling the sampling causes a significant increase in the variance of $N$, but the proportion of $N=0$ only slightly decreases. This indicates that the long-tail distribution is not mitigated. It is also evident that classes with extremely high sample counts at $20K$ sampling remain so at $40K$ sampling, demonstrating the inefficiency of blindly increasing the sampling size. In \textbf{(b)}, it can be observed that the bar chart obtained from FFHQ sampling is flatter than that of CelebA, indicating that using a more diverse image prior can effectively mitigate the long-tail distribution, specifically reflected in the smaller variance of $N$. In \textbf{(c)}, using a stronger GAN under the same image priors provides only minimal improvement for the long-tail distribution, as evidenced by a slight reduction in variance. Interestingly, classes that occupy more samples under the weaker GAN sampling often also appear in the strong GAN sampling, manifesting as symmetry in the bar charts above and below. However, charts derived from different prior samplings tend to be asymmetrical, as shown in \textbf{(b)}. This suggests that the mitigation of long-tail distributions by image priors primarily depends on the data distribution, rather than the advancement of the generative model.}}
\vspace{-12pt} 
\label{fig:logits}
\end{figure*}

\subsection{Data analysis on logits}

For training surrogate models \(M_s\), the typical paradigm involves using input-output pairs \((\mathcal{X}_{init},\mathcal{Y}_{init})\) derived from the \(M_t\). The \(M_t\) functions as the teacher model, and in a knowledge distillation manner~\cite{hinton2015distilling}, the information from the \(\mathcal{D}_{priv}\) is transferred to a locally pre-trained model which acts as the student model. This process is formalized as:
\begin{align}
\min_{M_s} \ \mathbb{E}_{(x, y) \sim (\mathcal{X}_{init}, \mathcal{Y}_{init})} \left[ D_{KL}(q(y|x) \| p(y|x; M_s)) \right],
\label{eq:KL}
\end{align}
where \(D_{KL}\) is Kullback-Leibler divergence~\cite{kullback1951information}, and the feature extractor of \(M_s\) is initialized from \(M_{pre}\), which is a recognition model pretrained on \(\mathcal{D}_{pre}\), a dataset whose distribution differs from \(\mathcal{D}_{priv}\).

However, we find that even with a large number (up to \(10K\)) of input-output pairs, the black-box MI attacker is still unable to effectively construct a high-quality surrogate model. The specific performance is the excessively low test accuracy of the \(M_s\) on the \(\mathcal{D}_{priv}\), as shown in ~\cref{table:experiment_loss_ablation}. This means that the resulting surrogate model obtained in this way is not sufficiently similar in predictive behavior to the target model, as the test accuracy serves as a crucial metric for the effect of knowledge distillation~\cite{gou2021knowledge} and model stealing~\cite{jagielski2020high,oliynyk2023know}. This further leads to the fact that performing white-box MIAs on the local substitution model can only obtain extremely poor attack results. Next, we analyze the causes of this low accuracy.

We use a VGGFace2 pre-trained ResNet50 model~\cite{VGG16} as the \(M_t\) and \(G\) trained on CelebA as the auxiliary image prior for our example. By analyzing \(20K\) logits, we find that the number of samples allocated to each class shows a very extreme long-tail distribution. Among the \(8631\) IDs, only \(9\%\) of the classes have no less than \(5\) samples, while \(65.9\%\) of the classes have no samples assigned at all. Notably, some classes occupy far more samples than the average, as shown in ~\cref{fig:logits}(a). We believe that the reason for this phenomenon is that CelebA, as an auxiliary dataset, has a significant difference from the target dataset VGGFace2 in the fine-grained distribution of facial feature combinations. Despite the extensive sample size, the intersection of the two covers only a portion of the private IDs. We further conjecture that blindly increasing the sample size has a limited contribution to improving the quality of surrogate models. This is because, for generative models, a large number of samples from different batches will have very close distributions despite the randomness. In simple terms, doubling the sample size will not substantially improve the extreme long-tail distribution. While there may be a slight increase in the number of intersecting IDs, the majority of the samples still fall into the already heavily populated classes, as shown in ~\cref{fig:logits}(a). More sampled data is redundant for the training of surrogate models, but greatly increases the query burden of black-box MIAs.

\vspace{-2pt}
One possible mitigation strategy involves using an auxiliary \(\mathcal{D}_{pub}\) that encompasses a richer combination of facial features, thereby ensuring greater diversity, to achieve more overlap with the private IDs. Ideally, when the auxiliary \(\mathcal{D}_{pub}\) is diverse enough and its data distribution overlaps with the \(\mathcal{D}_{priv}\) sufficiently, uniformly sampling from a high-quality generative model could effectively cover private IDs. Specifically, when utilizing the FFHQ pre-trained GAN model (FFHQ includes vastly more variation than CelebA~\cite{karras2019style}), with an equivalent sample size, FFHQ achieves a broader intersection and exhibits reduced variance in the number of data points per class. However, the issue of extreme long-tail distribution persists, as illustrated in ~\cref{fig:logits}(b). 
Even when using a more powerful GAN model, the mitigation of the long-tail problem remains limited, as shown in ~\cref{fig:logits}(c).
Considering that attackers are often unable to select the most suitable GAN model, our solution follows: Without relying on increasing sample sizes or enhancing auxiliary priors, within a limited sampling budget, we aim to extract as much information as possible from the private training data in the context of extreme long-tail distribution, to obtain higher-quality surrogate models.

\subsection{\large{SMILE}}
\vspace{-2pt}
Based on the analysis presented, we propose \textbf{S}urrogate \textbf{M}odel-based \textbf{I}nversion with \textbf{L}ong-tailed \textbf{E}nhancement (\textit{SMILE}), a two-step efficient black-box MIA. In the first step, we introduce a long-tailed surrogate training method specifically tailored for the MIA by integrating various long-tailed learning techniques. In the second step, using the results of attacking the local surrogate model as initialization, we apply a gradient-free optimization algorithm to enhance the effectiveness of the attack.

\begin{figure}[h]
  \centering
  \begin{overpic}[width=0.38\textwidth]{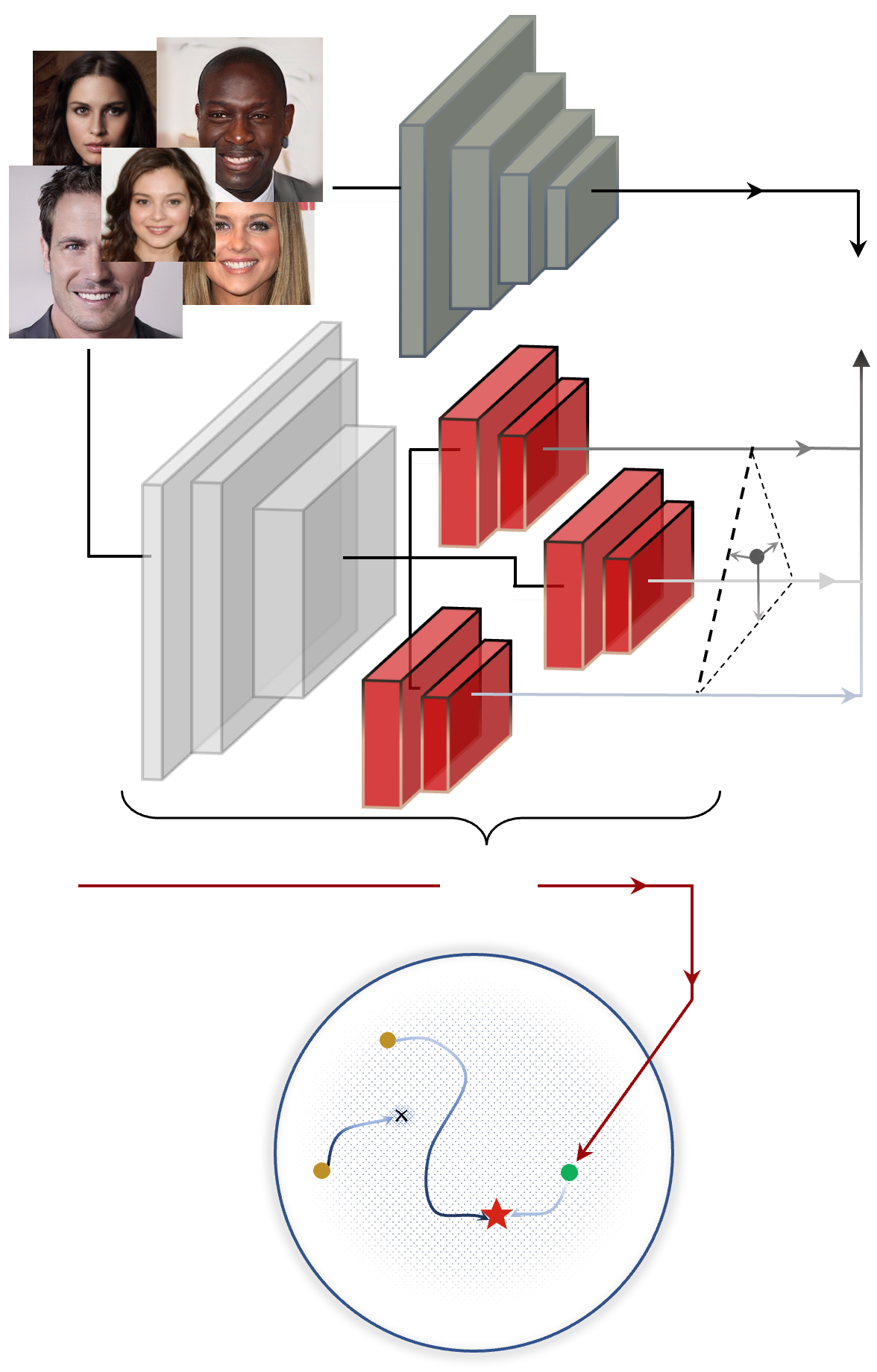}
  
  \put (48,89) {\large{\(M_t\)}}
  \put (45,79) {\small{\(\mathcal{L}_{KD}=D_{KL}+D_{CE}\)}}
  \put (53.1,50.3) {\footnotesize{\(\mathcal{L}_{\textit{Diversity}}\)}}
  \put (44,69) {\small{\(C_s^1\)}}
  \put (49,64) {\small{\(C_s^2\)}}
  \put (36,43) {\small{\(C_s^3\)}}
  \put (47.3,75.8) {\small{\textit{Top-k Reweighting}}}
  \put (4,98.5) {\textit{Initialization}}
  \put (4,50) {\large{\(E_\theta\)}}
  \put (33,34) {\large{\(M_s\)}}
  \put (47,45.3) {Step {\color[HTML]{9A0000}{\raisebox{-0.2ex}{\large \ding{172}}}}}
  \put (51.8,31.5) {Step {\color[HTML]{00008B}{\raisebox{-0.2ex}{\large \ding{173}}}}}

  \put (27,25.5) {\scriptsize{\(w_0\)}}
  \put (22.2,12.3) {\scriptsize{\(w_0\)}}
  \put (33.5,9.3) {\scriptsize{\(w_{\textit{target}}\)}}
  \put (42.2,12.8) {\scriptsize{\(w'\)}}

  \put (1,32) {\small{NGOpt}}
  \put (0,21){\includegraphics[scale=0.25]{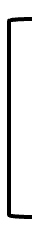}}
  \put (1.5,28.1) {\color[HTML]{00FF00}{\scriptsize{\textit{MetaModelOnePlusOne}}}}
  \put (1.8,25.8) {\tiny{CMA}}
  \put (1.88,23.8) {\tiny{PSO}}
  \put (2.0,22.6) {\scriptsize{\dots~\dots}}

  \end{overpic}
  \vspace{-5pt}
  \caption{The overall architecture of \textbf{\textit{SMILE}.}}
  \vspace{-10pt}
  \label{fig:smile-architecture}
\end{figure}

\begin{figure*}[]
\centering
\begin{overpic}[width=0.8\textwidth]{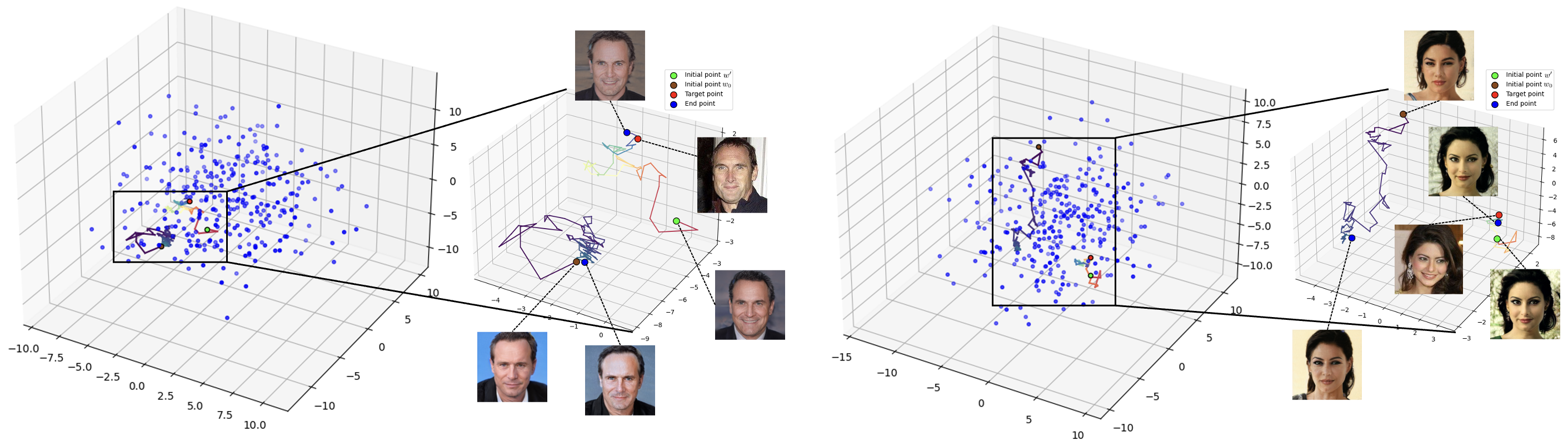}
\put (5,1.4) {\footnotesize{(a)}}
\put (56.5,1.4) {\footnotesize{(b)}}

\end{overpic}
\vspace{-8pt}
\caption{\footnotesize{\textbf{We visualized the optimization processes under different initializations.} $w_0$ is the sample with the highest confidence for the target ID in the sample pool, and $w'$ is the sample obtained from the white-box MIA on $M_s$. As shown in \textbf{(a)}, long-tailed surrogate training provides a more helpful initial point for black-box optimization, allowing it to approach the target point in very few iterative steps while avoiding local optima. \textbf{(b)} shows that long-tailed surrogate training fully captures the information of this ID, thereby providing a high-quality initial point close to the target point.}}
\vspace{-15pt}
\label{fig:pca_visualization}
\end{figure*}

\paragraph{Long-tailed surrogate training} Considering the attacker's access to only a very limited dataset of \((\mathcal{X}_{init}, \mathcal{Y}_{init})\) pairs, we update merely the parameters of the final two layers of \(M_s\) during training. Because fine-tuning \(M_{pre}\) with a too small dataset, which is misaligned with the private dataset, would lead to a catastrophic collapse in its feature extraction capabilities. ~\cref{sub:full_parameter} corroborates this view. To address the challenges posed by the extremely long-tail distribution of the training data, our strategy aims to maximize the extraction of information from the tail class 
while ensuring comprehensive learning of the head class 
information. Inspired by RIDE~\cite{wang2020long}, we design an ensemble method with Distribution-aware diversity loss~\cite{wang2020long}.

The pre-trained \(M_{pre}\) consists of a feature extractor \(f_\theta=\{f_{\theta 1},\dots,f_{\theta n}\}\) with \(n\) layers and a classifier \(C_{pre}\). For training \(M_s\), we freeze the parameters of the first \(n-1\) layers, using them as a shared backbone \(E_\theta=\{f_{\theta 1},\dots,f_{\theta n-1}\}\). 
We then construct \(N\) new ensemble classifiers \(C_s=\{f_{\theta n},C_t'\}\), where \(C_t'\) is randomly initialized but maintains the same output dimension as \(M_t\). The inference of \(M_s\) can be formalized as follows:
\begin{align}
M_s(x) = \frac{1}{N} \sum_{i=1}^N C_s^i (E_\theta(x)).
\label{eq:Ms}
\end{align}
All these \(N\) classifiers \(C_s\) are trained together using a distribution-aware diversity loss \(\mathcal{L}_{Diversity}\) and a distillation loss \(\mathcal{L}_{KD}\). The calculation of \(\mathcal{L}_{KD}\) is independent within the ensemble models, ensuring that each model complements the others~\cite{wang2020long}, as formalized below:
\begin{align}
\mathcal{L}_{KD}=\sum_{i=1}^N [D_{KL}(q(y|x)||p(y|x;C_s^i \circ E_\theta)) \label{eq:KD},\\
+~~\alpha_{ce} * D_{CE}(y_{pseudo},p(y|x;C_s^i \circ E_\theta))], \notag
\end{align}
where \((x,y) \in (\mathcal{X}_{init}, \mathcal{Y}_{init})\), \(\alpha_{ce}\) is the hyperparameter that balances \(D_{KL}\) with \(D_{CE}\). The ensemble comprises \(N\) models, set to 3 (ablation of \(N\) is in \cref{sub:ablation_number}). 
The pseudo-hard label \(y_{pseudo}\) is employed, where labels are built according to \(\mathcal{Y}_{init}\) by selecting the class with the highest probability. The guidance of pseudo-hard labels ensures efficient learning of head classes by the ensemble models.

\(\mathcal{L}_{Diversity}\) serves as a regularization term to promote complementary decisions from multiple models of the ensemble. The implementation strategy involves maximizing the Kullback-Leibler divergence between the prediction probabilities of each individual model and the ensemble's average prediction probabilities. To ensure training stability, we minimize the sum of the reciprocals as:
\begin{align}
\mathcal{L}_{Diversity}=\sum_{i=1}^N \frac{1}{D_{KL}(p_i(y|x;C_s^i \circ E_\theta)||p_{avg}(y|x;M_s))},
\label{eq:Diversity}
\end{align}
where \(p_i(y|x;C_s^i \circ E_\theta)\) denotes the predicted probability distribution of the \(i\)-th model.
To further mitigate the issue of long-tail distribution, we adopt a different approach from the common reweighting method~\cite{cao2019learning,cui2019class}. Considering that the Top-1 labels of the samples only cover a subset of all classes, and these classes are sparsely distributed across the entire set, we propose a \textit{Top-k Reweighting} strategy. We adjust the weights based on the frequency of each class appearing in the Top-k prediction probabilities, formalized as follows: For \(y_n \in \{y_1,\dots,y_{Step-1}\}\), find the class index of the corresponding Top-k largest logits \(T_n=\{t_{n1},\dots,t_{nk}\}\), where \(t_{nk}\) denotes the class index corresponding to the \(k\)-th largest logits of \(y_n\). For each class \(c \in \mathcal{C}_{target}=\{1, \ldots, N_{target}\}\), count the total number of its occurrences in the Top-k logits of all samples:
\begin{align}
\text{count}(c)=\sum_{n=1}^{N_{Step-1}} \sum_{j=1}^{k} \delta(t_{nj} = c)
,\delta(A) = 
\begin{cases} 
1, & if A\\
0, & else.
\end{cases} 
\label{eq:Top-k-count}
\end{align}
Set \(\beta=0.9\), and for each class \(c\), calculate the weights:
\begin{align}
\text{weights}(c) = \frac{1 - \beta}{1 - \beta^{\text{count}(c) + 1}}
\label{eq:Top-k-weight},\\
W = \text{weights}(t_{n1}).
\end{align}
In summary, the total loss for our proposed Long-tailed surrogate training can be described as follows:
\begin{align}
\mathcal{L}_{total}=W*\mathcal{L}_{KD} + \alpha_{diversity} * \mathcal{L}_{Diversity}.
\label{eq:total}
\end{align}
It is worth noting that although our total loss function has several hyperparameters, we empirically chose a fixed setting throughout our experiments (\(\alpha_{ce}=0.15,\alpha_{diversity}=10\), \textit{Top-10 Reweight}), which illustrates the robustness of our method. We perform Mirror-w~\cite{an2022mirror} on \(M_s\) and send the results \(w'\in\mathcal{W}_{init}'\) to the second step, formalized as:
\begin{align}
    \min_{w\in\mathcal{W}_{init}}\ \mathcal{L}_{id}(M_s(G(w)), c),
\label{eq:mirror_1}
\end{align}
\vspace{-18pt}
\begin{align}
    w_{i}=\text{\footnotesize{LeakyReLU}}
    (Clip(\text{\footnotesize{LeakyReLU}}^{-1}(w_i),\mu\pm\sigma)).
\label{eq:mirror_2}
\end{align}
where \(\mathcal{W}_{init}=G_{mapping}(\mathcal{Z}_{init})\). ~\cref{eq:mirror_2} serves as a regularization term, performing \(\mathcal{P}\) space pruning on \(w\) after each update. The upper and lower bounds are derived from the mean and variance of \(\mathcal{W}_{init}\) in \(\mathcal{P}\) space.

\vspace{-4pt}
\paragraph{Gradient-free black-box optimization} Considering the limitation that the initial sampling covers less than \(50\%\) of the private IDs. Although the results from the first step can be classified by \(M_s\) to the target ID, the corresponding facial features are often not accurately aligned with the target privacy instance. To improve the confidence of attack results in \(M_t\), we aim for more accurate feature matching and generalized attack outcomes. 

We employ the black-box optimization algorithm selector NGOpt. Given the latent feature dimension of \(512\) and a black-box query budget of \(2500\) (the reason for selecting this budget is presented in ~\cref{sub:why_2500}), NGOpt automatically determines the appropriate black-box optimization algorithm \textit{MetaModelOnePlusOne}~\cite{beyer2002evolution,rapin2018nevergrad}. We conduct black-box attacks on \(M_t\) with it, leveraging the initialization from the first step. 
The black-box optimization process is particularly challenging compared to the white-box setting due to the lack of gradient information, which makes it harder to search for private features solely based on the output.
To address this challenge, we further release the latent vector search capability of GANs. In the optimization process, we adopt the \(\mathcal{P}\) space clipping method~\cite{an2022mirror,zhu2020improved} and enlarge the clipping interval, formalized as follows:
\begin{align}
    \min_{w'\in\mathcal{W}_{init}'}\ \mathcal{L}_{id}(M_t(G(w')), c),
\label{eq:smile_1}
\end{align}
\vspace{-14pt}
\begin{align}
    w_{i}'=\text{\footnotesize{LeakyReLU}}(Clip(\text{\footnotesize{LeakyReLU}}^{-1}(w_i'),\mu\pm k * \sigma)).
\label{eq:smile_2}
\end{align}
After the optimization, the attacker obtains the final \(x^*=G(w^*)\). In the experiments, we set \(k=1.7\) empirically. The overall architecture of \textit{SMILE} is shown in ~\cref{fig:smile-architecture}. We further visualize the attack process of \textit{SMILE} by Principal Component Analysis (PCA) in ~\cref{fig:pca_visualization}.

\section{Experiments}
\label{sec:experiments}

In the experimental section, we evaluate the current SOTA MIAs targeting high-resolution scenarios, including Mirror~\cite{an2022mirror}, PPA~\cite{struppek2022plug}, and the SOTA black-box MIA RLBMI~\cite{han2023reinforcement}. We take the results of SOTA white-box MIAs as an upper bound for our attack's effectiveness, given that having access to the parameters significantly reduces the complexity of the optimization problem. We do not consider other MIAs targeting low-resolution scenarios, as their distortion in high-resolution scenarios is predictable.

\vspace{-2pt}
\subsection{Experimental setup}

For fairness, we adhered to the default configurations of other MIAs, including hyperparameters, GAN types, and the number of queries, as specified in their papers or official open-source repositories. Details are in \cref{supp-sec:experimental_setup_details}.

\vspace{-8pt}
\paragraph{Dataset.} We focused on facial recognition tasks, specifically the pre-trained face classification models for VGGFace~\cite{parkhi2015deep}, VGGFace2~\cite{cao2018vggface2}, and CASIA~\cite{yi2014learning} datasets. For image priors, \(\mathcal{D}_{pub}\) and \(\mathcal{D}_{priv}\) in all experiments originated from distinct distributions to maintain the practicality of MIA. Details are included in \cref{table:motivation_table_1} and \cref{supp-sec:experimental_setup_details}.

\begin{table*}[h!]
\centering
\resizebox{\textwidth}{!}{

\begin{tabular}{ccccccccccccc}
\toprule
\(\mathcal{D}_{priv}\)        & \multicolumn{4}{c}{VGGFace}                             & \multicolumn{8}{c}{CASIA}                                                                           \\ \hline
\(M_t\)           & \multicolumn{2}{c}{VGG16} & \multicolumn{2}{c}{VGG16BN} & \multicolumn{4}{c}{InceptionV1}                  & \multicolumn{4}{c}{SphereFace}                   \\ \hline
Method         & Acc@1$\uparrow$       & Acc@5$\uparrow$       & Acc@1$\uparrow$        & Acc@5$\uparrow$        & Acc@1$\uparrow$       & Acc@5$\uparrow$       & KNN Dist$\downarrow$ & Feat Dist$\downarrow$ & Acc@1$\uparrow$       & Acc@5$\uparrow$       & KNN Dist$\downarrow$ & Feat Dist$\downarrow$ \\ \hline
Mirror-w               & 81.63                              & 89.80                              & 77.55                              & 93.88                        & 57.14                              & 75.51                              & 446.96                        & 403.66                        & 61.22                              & 77.55                              & 379.51                        & 388.24                        \\
PPA                    & 97.96                              & 100.0                              & 100.0                              & 100.0                        & 77.55                              & 83.67                              & 385.72                        & 347.79                        & 57.14                              & 77.55                              & 363.42                        & 375.14                        \\ \hline
Mirror-b               & 59.18                              & 73.47                              & 57.14                              & 79.59                        & 28.57                              & 44.90                              & 601.762                       & 559.36                        & 16.33                              & 30.61                              & 425.63                        & 440.86                        \\
RLBMI                  & 71.43                              & 83.67                              & 71.43                              & {\color[HTML]{9A0000} 91.84} & 16.33                              & 36.73                              & 597.84                        & 556.45                        & 0.0                                & 6.12                               & 575.32                        & 578.50                        \\
\textbf{ResNet50*}               & 70.07$\pm$1.92                        & 87.76$\pm$1.67                        & 71.43$\pm$0.0                         & 78.23$\pm$1.92                  & 37.66$\pm$0.85                        & 53.41$\pm$1.26                        & 459.27                        & 419.53                        & 40.41$\pm$2.65                        & 58.89$\pm$3.36                        & 442.19                        & 452.48                        \\
\textbf{InceptionV1*}            & 71.43$\pm$1.67                        & {\color[HTML]{9A0000} 89.12$\pm$0.96} & 63.27$\pm$1.67                        & 72.79$\pm$0.96                  & 45.84$\pm$4.44                        & 63.00$\pm$0.38                        & {\color[HTML]{9A0000} 435.79} & {\color[HTML]{9A0000} 403.83} & 39.46$\pm$1.93                        & 58.50$\pm$0.96                        & 444.19                        & 445.67                        \\
\textbf{InceptionV3}            & {\color[HTML]{963400} 72.11$\pm$1.92} & 87.76$\pm$1.67                        & 64.63$\pm$0.96                        & 82.31$\pm$0.96                  & 43.54$\pm$2.54                        & {\color[HTML]{9A0000} 73.47$\pm$4.41} & 525.96                        & 477.83                        & {\color[HTML]{9A0000} 51.70$\pm$1.92} & 61.91$\pm$1.93                        & {\color[HTML]{9A0000} 392.27} & {\color[HTML]{9A0000} 404.40} \\
\textbf{MobileNetV2}            & 71.43$\pm$1.67                        & 82.99$\pm$1.93                        & 68.71$\pm$0.96                        & 85.03$\pm$1.93                  & {\color[HTML]{9A0000} 47.62$\pm$0.96} & 71.43$\pm$2.89                        & 624.43                        & 560.87                        & 48.30$\pm$2.54                        & {\color[HTML]{9A0000} 65.99$\pm$4.19} & 398.85                        & 407.20                        \\
\textbf{EfficientNetB0}         & 70.75$\pm$1.92                        & 84.35$\pm$2.55                        & 71.21$\pm$1.95                        & 79.45$\pm$1.5                   & 44.90$\pm$3.33                        & 65.99$\pm$0.96                        & 500.48                        & 458.35                        & 46.26$\pm$2.54                        & 63.26$\pm$2.89                        & 417.64                        & 424.17                        \\
\textbf{Swin-T}                 & 71.90$\pm$1.14                        & 83.53$\pm$1.84                        & {\color[HTML]{9A0000} 72.11$\pm$0.96} & 80.95$\pm$1.92                  & 45.58$\pm$0.96                        & 65.99$\pm$3.47                        & 496.31                        & 453.34                        & 48.30$\pm$2.54                        & 64.63$\pm$2.55                        & 396.13                        & 405.47                        \\ \bottomrule
\end{tabular}
}
\vspace{-6pt}
\label{table:experiment_celeba_VGGFace_CASIA_2500}
\end{table*}

\begin{table*}[]
\centering
\resizebox{\textwidth}{!}{

\begin{tabular}{cclccccccclcccccccc}
    \cline{1-10} \cline{12-19}
    \(\mathcal{D}_{priv}\) & \multicolumn{9}{c}{VGGFace2}                                                                                                                                                                                                                                                                &  & \multicolumn{8}{c}{VGGFace2}                                                                                                                                                                                                                                                        \\ \cline{1-10} \cline{12-19} 
    \(M_t\)                & \multicolumn{5}{c}{ResNet50}                                                                                                                          & \multicolumn{4}{c}{InceptionV1}                                                                                                     &  & \multicolumn{4}{c}{ResNet50}                                                                                                            & \multicolumn{4}{c}{InceptionV1}                                                                                                           \\ \cline{1-10} \cline{12-19} 
    Method                 & \multicolumn{2}{c}{Acc@1$\uparrow$}                              & Acc@5$\uparrow$                        & KNN Dist$\downarrow$                      & Feat Dist$\downarrow$                     & Acc@1$\uparrow$                              & Acc@5$\uparrow$                        & KNN Dist$\downarrow$                       & Feat Dist$\downarrow$                      &  & Acc@1$\uparrow$                              & Acc@5$\uparrow$                              & KNN Dist$\downarrow$                      & Feat Dist$\downarrow$                     & Acc@1$\uparrow$                              & Acc@5$\uparrow$                              & KNN Dist$\downarrow$                      & Feat Dist$\downarrow$                      \\ \cline{1-10} \cline{12-19} 
    Mirror-w               & \multicolumn{2}{c}{63.27}                              & 79.59                        & 313.33                        & 280.97                        & 51.02                              & 61.22                        & 2878.88                        & 2938.51                        &  & 81.63                              & 97.96                              & 227.06                        & 208.90                        & 81.63                              & 85.71                              & 2044.31                        & 2078.71                        \\
    PPA                    & \multicolumn{2}{c}{100.0}                              & 100.0                        & 161.54                        & 150.97                        & 93.88                              & 100.0                        & 1933.23                        & 2028.02                        &  & 97.96                              & 97.96                              & 149.35                        & 144.49                        & 100.0                              & 100.0                              & 1663.03                        & 1776.31                        \\ \cline{1-10} \cline{12-19} 
    Mirror-b               & \multicolumn{2}{c}{24.49}                              & 38.78                        & 369.55                        & 332.88                        & 26.53                              & 46.94                        & 2714.71                        & 2736.29                        &  & 44.90                              & 67.35                              & 289.38                        & 264.83                        & 55.10                              & 81.63                              & 1895.33                        & 1970.20                        \\
    RLBMI                  & \multicolumn{2}{c}{42.86}                              & {\color[HTML]{9A0000} 65.31} & 375.58                        & 328.12                        & 40.82                              & {\color[HTML]{9A0000} 65.31} & 2934.68                        & 2963.05                        &  & 63.27                              & 89.80                              & 323.07                        & 281.20                        & 71.43                              & 85.71                              & 2551.42                        & 2589.16                        \\
    \textbf{InceptionV1*}         & \multicolumn{2}{c}{{\color[HTML]{9A0000} 44.22$\pm$2.54}} & 61.90$\pm$5.36                  & {\color[HTML]{9A0000} 339.16} & {\color[HTML]{9A0000} 306.98} & {\color[HTML]{000000} 41.50$\pm$0.96} & 63.95$\pm$0.96                  & 2790.31                        & 2760.00                        &  & 68.71$\pm$1.92                        & {\color[HTML]{3531FF} 93.20$\pm$0.96} & 262.56                        & {\color[HTML]{3531FF} 237.75} & {\color[HTML]{3531FF} 89.80$\pm$1.67} & {\color[HTML]{3531FF} 94.56$\pm$0.96} & {\color[HTML]{3531FF} 1779.18} & 1899.43                        \\
    \textbf{EfficientNetB0}            & \multicolumn{2}{c}{39.46$\pm$0.96}                        & 49.66$\pm$1.92                  & 392.75                        & 351.45                        & {\color[HTML]{9A0000} 50.34$\pm$0.96}   & 63.27$\pm$1.67                  & {\color[HTML]{9A0000} 2667.81} & {\color[HTML]{9A0000} 2672.07} &  & {\color[HTML]{3531FF} 74.15$\pm$2.54} & 89.12$\pm$0.96                        & {\color[HTML]{3531FF} 257.65} & 239.22                        & 79.59$\pm$1.67                        & 89.80$\pm$0.0                         & 1805.03                        & {\color[HTML]{3531FF} 1882.68} \\ \cline{1-10} \cline{12-19} 
    \end{tabular}

}
\label{table:table:experiment_VGGFace2_2500}
\vspace{-6pt}
\end{table*}

\begin{table*}[h!]
\centering
\resizebox{\textwidth}{!}{

\begin{tabular}{ccccccccccccc}
    \toprule
    \(\mathcal{D}_{priv}\)        & \multicolumn{4}{c}{VGGFace}                                                                                                                       & \multicolumn{8}{c}{CASIA}                                                                                                                                                                                                                                                         \\ \hline
    \(M_t\)           & \multicolumn{2}{c}{VGG16}                                               & \multicolumn{2}{c}{VGG16BN}                                             & \multicolumn{4}{c}{InceptionV1}                                                                                                         & \multicolumn{4}{c}{SphereFace}                                                                                                          \\ \hline
    Method         & Acc@1$\uparrow$                              & Acc@5$\uparrow$                              & Acc@1$\uparrow$                           & Acc@5$\uparrow$                          & Acc@1$\uparrow$                              & Acc@5$\uparrow$                              & KNN Dist$\downarrow$                      & Feat Dist$\downarrow$                     & Acc@1$\uparrow$                              & Acc@5$\uparrow$                              & KNN Dist$\downarrow$                      & Feat Dist$\downarrow$                     \\ \hline
    Mirror-w       & 93.88                              & 100.0                              & 87.76                              & 100.0                              & 77.55                              & 87.76                              & 465.96                        & 415.81                        & 73.47                              & 91.84                              & 319.26                        & 330.82                        \\
    PPA            & 100.0                              & 100.0                              & 97.96                              & 100.0                              & 85.71                              & 95.92                              & 456.28                        & 404.90                        & 59.18                              & 83.67                              & 352.20                        & 361.18                        \\ \hline
    Mirror-b       & 59.18                              & 79.59                              & 61.22                              & 75.51                              & 46.94                              & 71.43                              & 693.57                        & 635.81                        & 22.45                             & 40.82                              & 393.98                        & 400.51                        \\
    RLBMI          & 67.35                              & 91.84                              & 61.22                              & 77.55                              & 53.06                              & 71.43                              & 598.79                        & 576.03                        & 16.33                             & 24.49                              & 520.43                        & 519.73                        \\
    \textbf{ResNet50*}       & 80.27$\pm$4.19                        & {\color[HTML]{3531FF} 91.16$\pm$1.92} & 56.46$\pm$2.54                        & 78.91$\pm$1.92                        & 49.36$\pm$5.46                        & 68.47$\pm$3.57                        & 575.16                        & 523.97                        & 38.39$\pm$4.53                        & 55.49$\pm$3.20                        & 406.14                        & 413.89                        \\
    \textbf{InceptionV1*}    & {\color[HTML]{3531FF} 80.95$\pm$3.47} & 91.16$\pm$0.96                        & 63.95$\pm$0.96                        & 80.95$\pm$0.96                        & 58.50$\pm$1.92                        & 80.27$\pm$0.96                        & {\color[HTML]{3531FF} 459.21} & {\color[HTML]{3531FF} 421.65} & 48.30$\pm$3.47                        & 65.99$\pm$2.54                        & 400.65                        & 410.08                        \\
    \textbf{InceptionV3}    & 72.79$\pm$1.92                        & 89.12$\pm$0.96                        & 62.59$\pm$0.97                        & 87.08$\pm$0.97                        & {\color[HTML]{3531FF} 64.63$\pm$2.55} & 80.95$\pm$2.54                        & 649.73                        & 581.97                        & 53.74$\pm$0.96                        & 67.35$\pm$1.66                        & 370.07                        & 376.73                        \\
    \textbf{MobileNetV2}    & 79.59$\pm$4.41                        & 90.48$\pm$0.96                        & {\color[HTML]{3531FF} 75.36$\pm$3.15} & {\color[HTML]{3531FF} 91.78$\pm$1.60} & 62.59$\pm$0.97                        & 76.87$\pm$2.54                        & 673.40                        & 607.23                        & {\color[HTML]{3531FF} 58.50$\pm$5.85} & {\color[HTML]{3531FF} 75.51$\pm$1.67} & {\color[HTML]{3531FF} 346.74} & 362.32                        \\
    \textbf{EfficientNetB0} & 72.79$\pm$2.54                        & 89.80$\pm$1.66                        & 70.75$\pm$3.85                        & 80.95$\pm$3.47                        & 59.86$\pm$2.55                        & 73.47$\pm$3.33                        & 590.52                        & 534.32                        & 55.78$\pm$1.92                        & 70.07$\pm$0.96                        & 356.78                        & {\color[HTML]{3531FF} 359.19} \\
    \textbf{Swin-T}         & 80.95$\pm$1.92                        & 88.44$\pm$1.93                        & 68.71$\pm$0.96                        & 85.71$\pm$1.67                        & 59.58$\pm$1.76                        & {\color[HTML]{3531FF} 87.01$\pm$4.12} & 578.46                        & 515.81                        & 47.94$\pm$0.83                        & 72.60$\pm$0.86                        & 362.93                        & 374.42                        \\ \bottomrule
    \end{tabular}

}


\vspace{-4pt}
\caption{\footnotesize{\textbf{Performance report of MIAs on different private datasets and target model architectures.} The {\color[HTML]{9A0000}red} / {\color[HTML]{3531FF}blue} indicates the optimal \textbf{black-box} MIA results when the image prior is {\color[HTML]{9A0000}CelebA} / {\color[HTML]{3531FF}FFHQ}. The bolded \textbf{Arch} indicates the surrogate model architecture used by \textbf{\textit{SMILE}}. \textbf{*} refers to the surrogate model initialized with a pre-trained face recognition model on the Internet. \textbf{Swin-T} is the abbreviation for \textbf{Swin Transformer}.}}
\vspace{-15pt}
\label{table:experiment_ffhq_VGGFace_CASIA_2500}
\end{table*}

\vspace{-8pt}
\paragraph{Models.} Following Mirror~\cite{an2022mirror}, we utilize GANs~\cite{genforce2020} pre-trained on CelebA~\cite{liu2015deep} and FFHQ~\cite{karras2019style}. For the setting of the pre-trained model that initializes the surrogate model, we ensure that their training data come from distributions different from the private training data. Specifically, for VGGFace and CASIA, we use pre-trained ResNet50~\cite{VGG16} and InceptionV1~\cite{INCEPTIONv1} from VGGFace2; for VGGFace2, InceptionV1~\cite{INCEPTIONv1} pre-trained on CASIA is used. To verify the robustness of the proposed long-tailed surrogate training method for \(M_{pre}\), we self-train models across multiple architectures~\cite{liu2022swin,sandler2018mobilenetv2,szegedy2016rethinking,tan2019efficientnet}, details are in \cref{supp-sec:experimental_setup_details}.

\vspace{-12pt}
\paragraph{Evaluation metrics.} Following previous research~\cite{kahla2022label,zhang2020secret,nguyen2023re}, we select three metrics: Attack Accuracy (\textbf{Attack Acc}) : Following Mirror~\cite{an2022mirror}, we employ two models pre-trained on the same dataset, each serving as the evaluation model for the other, and report the \textbf{Acc@1} and \textbf{Acc@5}; K-Nearest Neighbors Distance (\textbf{KNN Dist}); Feature Distance (\textbf{Feat Dist}), details are in \cref{supp-sec:experimental_setup_details}.

\subsection{Experimental results}

\begin{figure}[h!]
\centering
\vspace{0.6cm}
\begin{overpic}[width=0.47\textwidth]{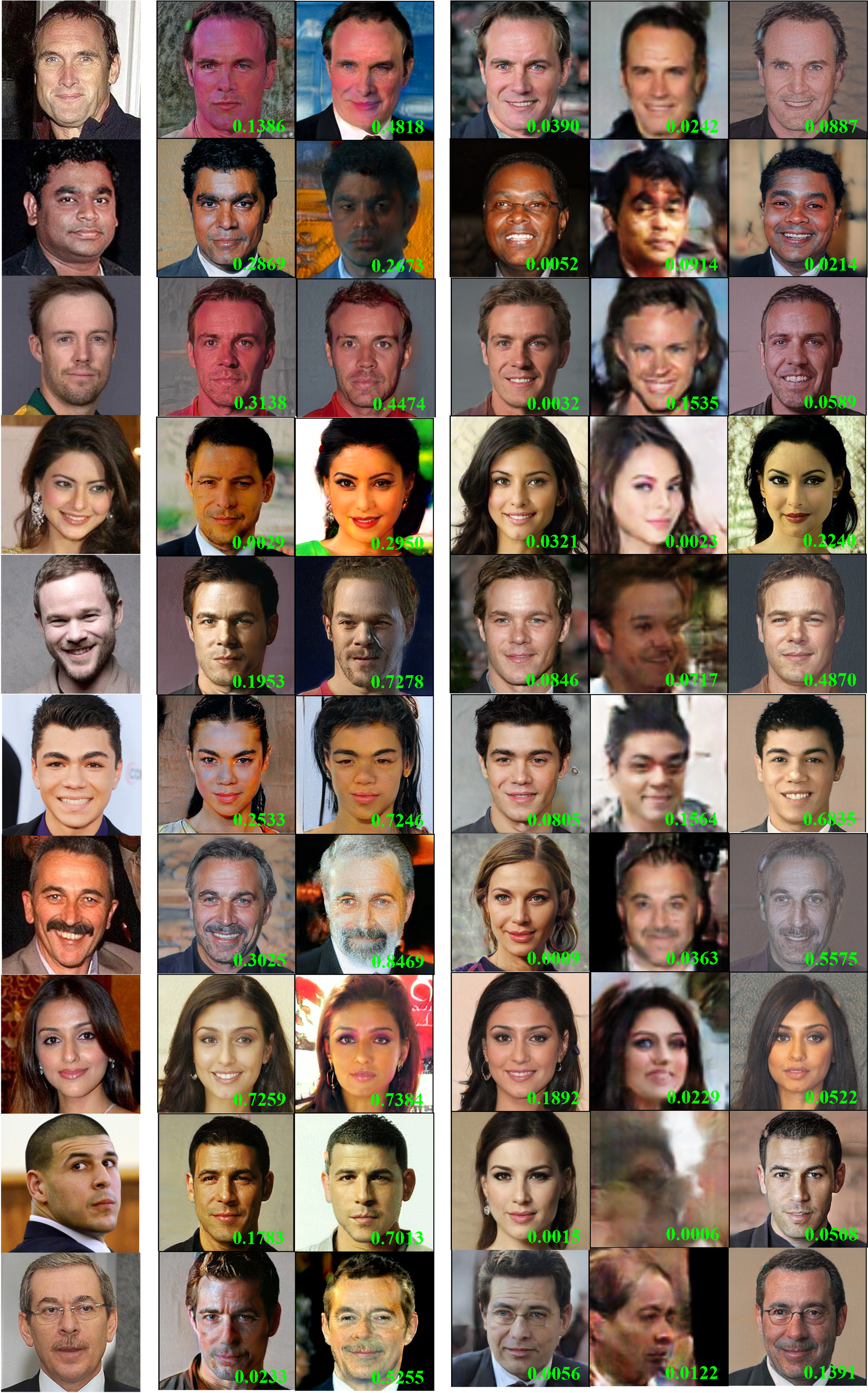}

  \put (2.1,102) {\scriptsize{Ground}}
  \put (2.85,100.4) {\scriptsize{Truth}}

  \put (12.8,100.4) {\scriptsize{Mirror-w}}
  \put (24.5,100.4) {\scriptsize{PPA}}

  \put (33.9,100.4) {\scriptsize{Mirror-b}}
  \put (44.3,100.4) {\scriptsize{RLBMI}}
  \put (54.4,100.4) {\scriptsize{SMILE}}

  \end{overpic}
  \vspace{-4pt}
  \caption{\footnotesize{\textbf{Results of MIAs.} \(M_t\) is ResNet50 and CelebA for image prior. The \(M_s\) of \textit{SMILE} is InceptionV1.}}
  \vspace{-20pt}
  \label{fig:attack_result}
\end{figure}
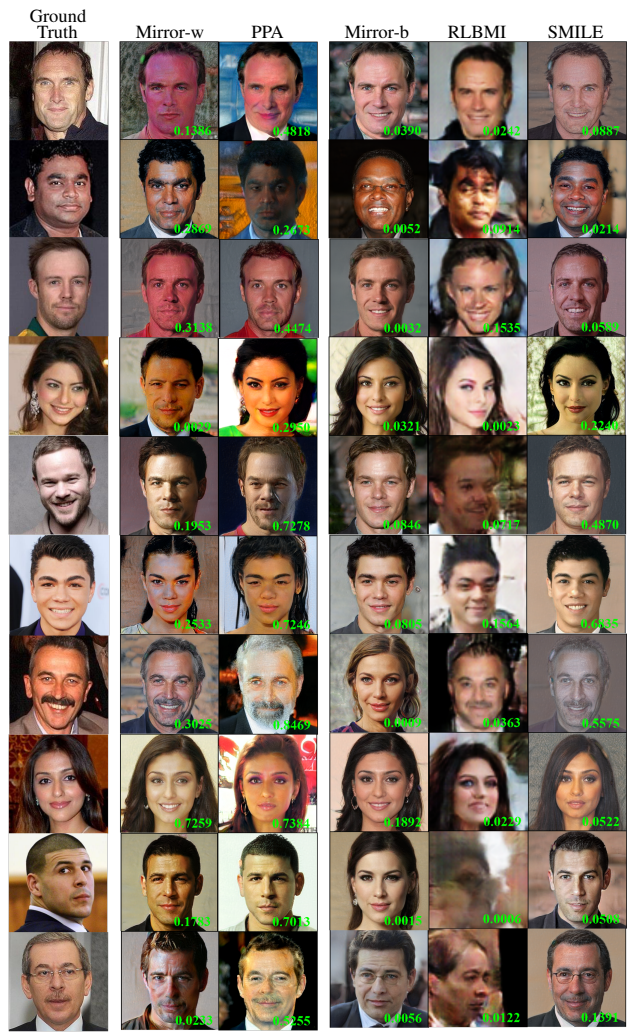

\cref{table:experiment_ffhq_VGGFace_CASIA_2500} presents a comprehensive overview of \textit{SMILE}'s attack results, illustrating that our method is SOTA for black-box MIA under most conditions, spanning various private datasets and diverse targeted pretrained models. The employment of multiple surrogate model architectures and image priors underscores \textit{SMILE}'s robustness, qualitative results are in \cref{fig:attack_result} and \cref{supp-sec:qualitative_details}. 
It can be observed that the effectiveness of \textit{SMILE} varies with different surrogate model architectures. This variation is due to different surrogate models' capability to extract features and their alignment with private datasets, affecting their adaptation to the target domain and consequently impacting the long-tailed surrogate training. 
To verify the robustness of our method, we fixed a uniform set of hyperparameters in the training of surrogate models, resulting in certain quality differences among different architectures. We consider this acceptable.

\begin{figure}[h!]
  \centering
  \vspace{-2pt}
  \begin{overpic}[width=0.45\textwidth]{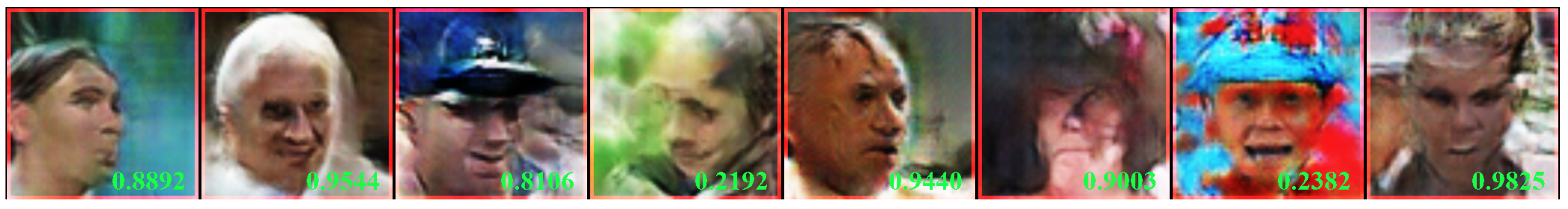}

  \end{overpic}
  \vspace{-8pt}
  \caption{\footnotesize{\textbf{Some results of RLBMI on VGG.} The image is seriously distorted, but the evaluation model has high confidence in them.}}
  \label{fig:RLBMI_inflated_metrics.png}
\end{figure}

For RLBMI, which performs suboptimally, we observe inflated metrics. This inflation occurs because under a large number of queries, some results are optimized into adversarial samples with transferability. Consequently, certain outcomes from RLBMI appear significantly distorted, yet they receive high confidence scores from the evaluation model, as shown in \cref{fig:RLBMI_inflated_metrics.png}. This phenomenon is more obvious when targeting the VGG dataset, where a relatively small number of classes are easier to generate adversarial examples. In contrast, \(\mathcal{P}\) space clipping effectively mitigates this problem.

\begin{table*}[h!]
\centering
\resizebox{\textwidth}{!}{

\begin{tabular}{cclccccccclcccccccc}
    \cline{1-10} \cline{12-19}
    {\color[HTML]{000000} \(\mathcal{D}_{priv}\)} &
      \multicolumn{9}{c}{{\color[HTML]{000000} VGGFace2, InceptionV1}} &
      {\color[HTML]{000000} } &
      \multicolumn{8}{c}{{\color[HTML]{000000} VGGFace2, ResNet50}} \\ \cline{1-10} \cline{12-19} 
    {\color[HTML]{000000} \(\mathcal{D}_{pub}\)} &
      \multicolumn{5}{c}{{\color[HTML]{000000} CelebA}} &
      \multicolumn{4}{c}{{\color[HTML]{000000} FFHQ}} &
      {\color[HTML]{000000} } &
      \multicolumn{4}{c}{{\color[HTML]{000000} CelebA}} &
      \multicolumn{4}{c}{{\color[HTML]{000000} FFHQ}} \\ \cline{1-10} \cline{12-19} 
    {\color[HTML]{000000} Method} &
      \multicolumn{2}{c}{{\color[HTML]{000000} \begin{tabular}[c]{@{}c@{}}\scriptsize{\textit{Base}}\\[-0.1cm] \scriptsize{$D_{KL}$}\end{tabular}}} &
      {\color[HTML]{000000} \begin{tabular}[c]{@{}c@{}}\scriptsize{\textbf{\textit{SMILE}}}\\[-0.1cm] \scriptsize{$D_{KL}$+$D_{Diversity}$}\end{tabular}} &
      {\color[HTML]{000000} \begin{tabular}[c]{@{}c@{}}\scriptsize{\textbf{\textit{SMILE}}}\\[-0.1cm] \scriptsize{··· +$D_{CE}$}\end{tabular}} &
      {\color[HTML]{000000} \begin{tabular}[c]{@{}c@{}}\scriptsize{\textbf{\textit{SMILE}}}\\[-0.1cm] \scriptsize{··· +\textit{Top-k Reweighting}}\end{tabular}} &
      {\color[HTML]{000000} \begin{tabular}[c]{@{}c@{}}\scriptsize{\textit{Base}}\\[-0.1cm] \scriptsize{$D_{KL}$}\end{tabular}} &
      {\color[HTML]{000000} \begin{tabular}[c]{@{}c@{}}\scriptsize{\textbf{\textit{SMILE}}}\\[-0.1cm] \scriptsize{$D_{KL}$+$D_{Diversity}$}\end{tabular}} &
      {\color[HTML]{000000} \begin{tabular}[c]{@{}c@{}}\scriptsize{\textbf{\textit{SMILE}}}\\[-0.1cm] \scriptsize{··· +$D_{CE}$}\end{tabular}} &
      {\color[HTML]{000000} \begin{tabular}[c]{@{}c@{}}\scriptsize{\textbf{\textit{SMILE}}}\\[-0.1cm] \scriptsize{··· +\textit{Top-k Reweighting}}\end{tabular}} &
      {\color[HTML]{000000} } &
      {\color[HTML]{000000} \begin{tabular}[c]{@{}c@{}}\scriptsize{\textit{Base}}\\[-0.1cm] \scriptsize{$D_{KL}$}\end{tabular}} &
      {\color[HTML]{000000} \begin{tabular}[c]{@{}c@{}}\scriptsize{\textbf{\textit{SMILE}}}\\[-0.1cm] \scriptsize{$D_{KL}$+$D_{Diversity}$}\end{tabular}} &
      {\color[HTML]{000000} \begin{tabular}[c]{@{}c@{}}\scriptsize{\textbf{\textit{SMILE}}}\\[-0.1cm] \scriptsize{··· +$D_{CE}$}\end{tabular}} &
      {\color[HTML]{000000} \begin{tabular}[c]{@{}c@{}}\scriptsize{\textbf{\textit{SMILE}}}\\[-0.1cm] \scriptsize{··· +\textit{Top-k Reweighting}}\end{tabular}} &
      {\color[HTML]{000000} \begin{tabular}[c]{@{}c@{}}\scriptsize{\textit{Base}}\\[-0.1cm] \scriptsize{$D_{KL}$}\end{tabular}} &
      {\color[HTML]{000000} \begin{tabular}[c]{@{}c@{}}\scriptsize{\textbf{\textit{SMILE}}}\\[-0.1cm] \scriptsize{$D_{KL}$+$D_{Diversity}$}\end{tabular}} &
      {\color[HTML]{000000} \begin{tabular}[c]{@{}c@{}}\scriptsize{\textbf{\textit{SMILE}}}\\[-0.1cm] \scriptsize{··· +$D_{CE}$}\end{tabular}} &
      {\color[HTML]{000000} \begin{tabular}[c]{@{}c@{}}\scriptsize{\textbf{\textit{SMILE}}}\\[-0.1cm] \scriptsize{··· +\textit{Top-k Reweighting}}\end{tabular}} \\ \cline{1-10} \cline{12-19} 
    {\color[HTML]{FD6864} \textbf{InceptionV1*}} &
      \multicolumn{2}{c}{{\color[HTML]{000000} 5.00/11.74}} &
      {\color[HTML]{32CB00} 12.94/25.11} &
      {\color[HTML]{32CB00} 16.21/29.59} &
      {\color[HTML]{32CB00} 21.84/39.29} &
      {\color[HTML]{000000} 12.07/26.20} &
      {\color[HTML]{32CB00} 25.42/43.99} &
      {\color[HTML]{32CB00} 28.67/47.60} &
      {\color[HTML]{32CB00} 36.54/58.90} &
      {\color[HTML]{32CB00} } &
      {\color[HTML]{000000} 2.58/6.88} &
      {\color[HTML]{32CB00} 5.25/11.92} &
      {\color[HTML]{32CB00} 6.18/14.40} &
      {\color[HTML]{32CB00} 8.91/19.40} &
      {\color[HTML]{000000} 6.78/16.74} &
      {\color[HTML]{32CB00} 12.85/26.91} &
      {\color[HTML]{32CB00} 14.00/27.86} &
      {\color[HTML]{32CB00} 20.45/38.98} \\
    {\color[HTML]{FD6864} \textbf{EfficientNetB0}} &
      \multicolumn{2}{c}{{\color[HTML]{000000} 4.32/11.51}} &
      {\color[HTML]{32CB00} 6.47/14.90} &
      {\color[HTML]{32CB00} 6.85/15.40} &
      {\color[HTML]{32CB00} 9.80/20.74} &
      {\color[HTML]{000000} 7.77/18.61} &
      {\color[HTML]{32CB00} 11.57/24.67} &
      {\color[HTML]{32CB00} 11.66/}{\color[HTML]{CB0000}24.06} &
      {\color[HTML]{32CB00} 17.46/33.22} &
      {\color[HTML]{32CB00} } &
      {\color[HTML]{000000} 2.03/5.82} &
      {\color[HTML]{32CB00} 2.49/6.69} &
      {\color[HTML]{32CB00} 2.51/}{\color[HTML]{CB0000}6.56} &
      {\color[HTML]{32CB00} 3.08/7.76} &
      {\color[HTML]{000000} 4.05/11.44} &
      {\color[HTML]{32CB00} 4.87/12.81} &
      {\color[HTML]{32CB00} 5.01/}{\color[HTML]{CB0000}12.69} &
      {\color[HTML]{32CB00} 7.01/16.55} \\
    {\color[HTML]{CB0000} \textbf{InceptionV1*}} &
      \multicolumn{2}{c}{{\color[HTML]{000000} 6.36/14.77}} &
      {\color[HTML]{32CB00} 10.37/21.9} &
      {\color[HTML]{32CB00} 16.31/30.94} &
      {\color[HTML]{32CB00} 20.63/37.44} &
      {\color[HTML]{000000} 15.39/31.01} &
      {\color[HTML]{32CB00} 24.13/44.07} &
      {\color[HTML]{32CB00} 25.96/44.97} &
      {\color[HTML]{32CB00} 40.17/61.00} &
      {\color[HTML]{32CB00} } &
      {\color[HTML]{000000} 3.48/8.64} &
      {\color[HTML]{32CB00} 4.84/11.15} &
      {\color[HTML]{32CB00} 5.76/13.06} &
      {\color[HTML]{32CB00} 9.16/19.67} &
      {\color[HTML]{000000} 9.34/21.16} &
      {\color[HTML]{32CB00} 14.35/30.17} &
      {\color[HTML]{32CB00} 17.84/34.12} &
      {\color[HTML]{32CB00} 23.78/42.76} \\
    {\color[HTML]{CB0000} \textbf{EfficientNetB0}} &
      \multicolumn{2}{c}{{\color[HTML]{000000} 6.19/15.18}} &
      {\color[HTML]{32CB00} 9.46/20.67} &
      {\color[HTML]{32CB00} 10.15/21.01} &
      {\color[HTML]{32CB00} 15.45/29.81} &
      {\color[HTML]{000000} 9.91/22.63} &
      {\color[HTML]{32CB00} 16.40/33.05} &
      {\color[HTML]{CB0000} 16.15/31.60} &
      {\color[HTML]{32CB00} 24.06/42.86} &
      {\color[HTML]{32CB00} } &
      {\color[HTML]{000000} 2.97/8.15} &
      {\color[HTML]{32CB00} 3.67/9.03} &
      {\color[HTML]{32CB00} 3.71/9.28} &
      {\color[HTML]{32CB00} 4.75/11.59} &
      {\color[HTML]{000000} 5.67/15.45} &
      {\color[HTML]{32CB00} 7.40/17.47} &
      {\color[HTML]{32CB00} 7.78/18.06} &
      {\color[HTML]{32CB00} 11.57/25.07} \\
    {\color[HTML]{680100} \textbf{InceptionV1*}} &
      \multicolumn{2}{c}{{\color[HTML]{000000} 8.53/18.21}} &
      {\color[HTML]{32CB00} 19.28/35.47} &
      {\color[HTML]{32CB00} 22.78/39.34} &
      {\color[HTML]{32CB00} 24.87/42.08} &
      {\color[HTML]{000000} 19.14/36.43} &
      {\color[HTML]{32CB00} 36.51/57.61} &
      {\color[HTML]{32CB00} 40.38/60.68} &
      {\color[HTML]{32CB00} 45.12/64.68} &
      {\color[HTML]{32CB00} } &
      {\color[HTML]{000000} 5.18/11.70} &
      {\color[HTML]{32CB00} 8.88/19.06} &
      {\color[HTML]{32CB00} 11.45/23.31} &
      {\color[HTML]{32CB00} 14.27/28.28} &
      {\color[HTML]{000000} 12.73/26.19} &
      {\color[HTML]{32CB00} 24.64/44.80} &
      {\color[HTML]{32CB00} 24.79/}{\color[HTML]{CB0000}43.80} &
      {\color[HTML]{32CB00} 32.29/52.51} \\
    {\color[HTML]{680100} \textbf{EfficientNetB0}} &
      \multicolumn{2}{c}{{\color[HTML]{000000} 7.91/18.60}} &
      {\color[HTML]{32CB00} 13.56/27.98} &
      {\color[HTML]{32CB00} 16.49/31.14} &
      {\color[HTML]{32CB00} 20.91/37.94} &
      {\color[HTML]{000000} 12.80/27.33} &
      {\color[HTML]{32CB00} 25.39/44.90} &
      {\color[HTML]{CB0000} 23.86/42.78} &
      {\color[HTML]{32CB00} 31.54/51.59} &
      {\color[HTML]{32CB00} } &
      {\color[HTML]{000000} 4.14/10.83} &
      {\color[HTML]{32CB00} 5.19/12.51} &
      {\color[HTML]{32CB00} 6.25/14.62} &
      {\color[HTML]{32CB00} 7.13/16.81} &
      {\color[HTML]{000000} 8.45/19.87} &
      {\color[HTML]{32CB00} 12.50/27.03} &
      {\color[HTML]{32CB00} 13.87/28.61} &
      {\color[HTML]{32CB00} 17.51/34.26} \\ \cline{1-10} \cline{12-19} 
    \end{tabular}

}
\vspace{-6pt}
\caption{\footnotesize{\textbf{Ablation study of loss terms.} We set the sample sizes to {\color[HTML]{FD6864}$2.5K$}, {\color[HTML]{CB0000}$5K$}, and {\color[HTML]{680100}$10K$}. For each setting, the loss terms are added sequentially from left to right, with {\color[HTML]{32CB00}green} indicating performance improvement and {\color[HTML]{CB0000}red} indicating performance decline. It can be observed that the performance of the surrogate models generally shows an upward trend and is significantly better than \textit{Base}. This suggests that each loss term contributes.}}
\label{table:experiment_loss_ablation}
\end{table*}

\begin{figure*}[h!]
\centering
\begin{overpic}[width=0.99\textwidth]{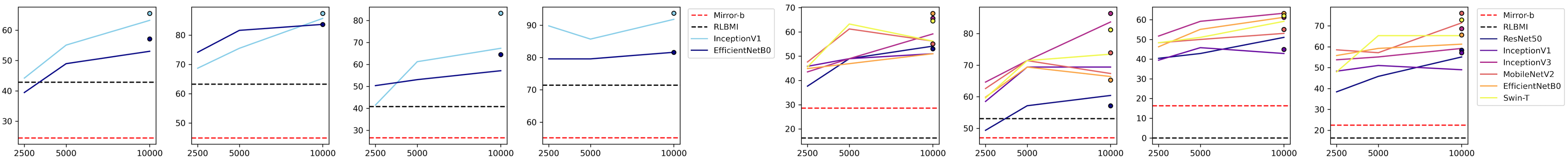}

  \put (1.2,10.3) {\scriptsize{Acc@1}}
  
  \put (1.1,-1.8) {\scriptsize{Number of queries}}
  \put (6.8,10.4) {\tiny{VGGFace2, ResNet50}}
  \put (4,-0.5) {\tiny{CelebA}}
  \put (15.2,-0.5) {\tiny{FFHQ}}

  \put (28.2,10.4) {\tiny{VGGFace2, InceptionV1}}
  \put (26.4,-0.5) {\tiny{CelebA}}
  \put (37.7,-0.5) {\tiny{FFHQ}}

  \put (56.8,10.4) {\tiny{CASIA, InceptionV1}}
  \put (54,-0.5) {\tiny{CelebA}}
  \put (65.4,-0.5) {\tiny{FFHQ}}

  \put (79.4,10.4) {\tiny{CASIA, SphereFace}}
  \put (76.3,-0.5) {\tiny{CelebA}}
  \put (87.8,-0.5) {\tiny{FFHQ}}




  
  \end{overpic}
  \caption{\footnotesize{\textbf{Setting initial sample sizes to $2.5K$, $5K$, and $10K$ reveals a gradual increase in the effectiveness of \textit{SMILE}.} This trend suggests that while a larger sample size does not directly address the long-tail distribution, it provides more privacy information. The surrogate model, relying on more information, can offer higher quality initial points, making the subsequent optimization process easier. Additionally, by extending the optimization steps from $1K$ to $2K$, an enhancement in efficacy is observable (the \textbf{dots} in the graph). This demonstrates \textit{SMILE}'s potential when expanding the query budget.}}
  \vspace{-14pt}
  \label{fig:line_chart}
\end{figure*}
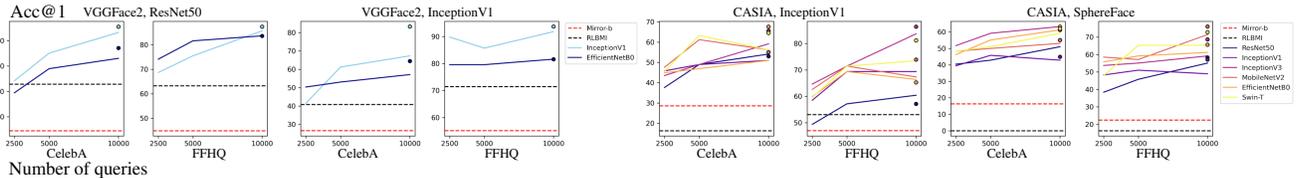

\vspace{-14pt}
\paragraph{Ablation study.} We incrementally add loss terms and observe their impact on the performance of the surrogate model. The \textit{Base} is the model obtained by applying \(D_{KL}\) and only updating the final classifier layer. \cref{table:experiment_loss_ablation} shows that in various settings, the increase in the loss terms of \textit{SMILE} almost invariably leads to positive effects, and the quality of the obtained surrogate models significantly surpasses that of \textit{Base}. We further increase the initial sampling size. \cref{fig:line_chart} indicates that this enhancement improves \textit{SMILE}'s performance, highlighting its potential for large-scale query.

\vspace{-10pt}
\paragraph{Defense.} We further evaluate various defense mechanisms, including BiDO~\cite{peng2022bilateral}, MID~\cite{wang2021improving}, LS~\cite{struppek2023careful}, and TL~\cite{ho2024model}. As shown in ~\cref{table:experiment_defense}, although there is a decrease in performance, our method still significantly outperforms existing black-box MIAs. We also find that defenses have a much greater effect on black-box MIAs than on white-box MIAs. We will focus on developing black-box MIAs that are insensitive to defenses as our future research interest. Details are in ~\cref{supp-sec:defenses_details}.

\vspace{-10pt}
\paragraph{Other experiments.} We further use art face~\cite{karras2020training} as image priors, which have a significant distribution difference from private data, as shown in ~\cref{sub:art_face}. We discuss the challenges faced by label-only MIA under pre-trained models and large-scale private ID settings, and introduce the concept of Attack-sensitive ID, which includes General Attack-sensitive ID and Dataset-specific Attack-sensitive ID. Details are provided in \cref{sub:challenge_label_only}.

\begin{table}[h!]
\centering
\resizebox{0.46\textwidth}{!}{
\begin{tabular}{ccccccccccc}
\toprule
                       &                                    &       & \multicolumn{2}{c}{PPA} & \multicolumn{2}{c}{Mirror-b} & \multicolumn{2}{c}{RLBMI} & \multicolumn{2}{c}{SMILE} \\ \hline
Defenses               & Hyperparameters                    & Acc   & Acc@1$\uparrow$      & Acc@5$\uparrow$      & Acc@1$\uparrow$         & Acc@5$\uparrow$        & Acc@1$\uparrow$       & Acc@5$\uparrow$       & Acc@1$\uparrow$       & Acc@5$\uparrow$       \\ \hline
                       & {\color[HTML]{9A0000} 0.006, 0.06} & 92.20 (2.27$\downarrow$) & 93.88      & 95.92      & 12.24         & 16.33        & 8.16        & 16.32       & 22.45       & 40.82       \\
\multirow{-2}{*}{BiDO} & {\color[HTML]{00009B} 0.03, 0.3}   & 91.57 (1.63$\downarrow$) & 55.11      & 75.51      & 6.12          & 14.29        & 8.16        & 22.44       & 22.49       & 36.73       \\ \hline
                       & {\color[HTML]{9A0000} 0.005}       & 91.31 (3.16$\downarrow$) & 93.88      & 100.00     & 8.16          & 24.49        & 6.12        & 22.44       & 14.29       & 26.53       \\
\multirow{-2}{*}{MID}  & {\color[HTML]{00009B} 0.005}       & 89.50 (3.70$\downarrow$)  & 83.67      & 93.87      & 4.08          & 4.08         & 18.36       & 24.48       & 6.12        & 10.20       \\ \hline
                       & {\color[HTML]{9A0000} -0.001}      & 92.40 (2.07$\downarrow$) & 87.76      & 95.92      & 8.16          & 10.20        & 4.08        & 14.28       & 30.61       & 38.78       \\
\multirow{-2}{*}{LS}   & {\color[HTML]{00009B} -0.0005}     & 92.48 (0.72$\downarrow$) & 65.31      & 77.55      & 10.20         & 14.29        & 8.16        & 12.24       & 24.49       & 34.69       \\ \hline
                       & {\color[HTML]{9A0000} Block 4}     & 93.82 (0.65$\downarrow$) & 81.63      & 95.92      & 6.12          & 10.20        & 14.28       & 16.32       & 20.41       & 36.73       \\
\multirow{-2}{*}{TL}   & {\color[HTML]{00009B} Block 3}     & 91.98 (1.22$\downarrow$) & 34.69      & 59.18      & 2.04          & 6.12         & 4.08        & 6.12        & 10.20       & 24.49       \\ 
\bottomrule
\end{tabular}
}
\vspace{-4pt}
\caption{\footnotesize{\textbf{Performance of MIAs under defenses.} The \(\mathcal{D}_{priv}\) is VGGFace2. {\color[HTML]{9A0000}Red} refers to {\color[HTML]{9A0000}MobileNetV2}, and {\color[HTML]{00009B}blue} refers to {\color[HTML]{00009B}Swin Transformer.}}}
\label{table:experiment_defense}
\end{table}

\vspace{-18pt}
\section{Conclusion}
\label{sec:conclusion}
\vspace{-4pt}

In this paper, we introduce \textit{SMILE}, an efficient black-box MIA. By combining long-tailed surrogate training and gradient-free black-box optimization, \textit{SMILE} outperforms existing black-box MIAs while requiring about 5\% of the query overhead. Experiments on various datasets and model architectures demonstrate its robustness.

\newpage

\clearpage

\section*{Acknowledgements}
We thank Xinli Yue for the discussion regarding long-tailed learning. This work was supported by the Key R\&D Project of Hubei Province under Grant No.2023BAB165. This work was supported by Ant Group. 

{
    \small
    \bibliographystyle{ieeenat_fullname}
    \bibliography{main}
}

\clearpage
\setcounter{page}{1}
\maketitlesupplementary

\appendix
 
\setcounter{figure}{0} 
\setcounter{table}{0} 
\setcounter{section}{0} 

\section{Additional experimental results}
\label{supp-sec:additional_results}

\subsection{Full parameter fine-tuning weakens the extraction ability}
\label{sub:full_parameter}

\vspace{-6pt}
\begin{table}[h!]
\centering
\resizebox{.47\textwidth}{!}{
\begin{tabular}{ccccc}
\hline
\(M_t\)         & \multicolumn{4}{c}{VGGFce2, ResNet50}                 \\ \hline
Image Priors & \multicolumn{2}{c}{CelebA} & \multicolumn{2}{c}{FFHQ} \\ \hline
Method       & \textit{Base}         & $\textit{Base}_{full}$  & \textit{Base}        & $\textit{Base}_{full}$ \\ \hline
{\color[HTML]{FD6864}InceptionV1*}  & 2.58/6.88    & 0.42/1.47   & 6.78/16.74  & 1.09/3.48  \\
{\color[HTML]{CB0000}InceptionV1*}  & 3.48/8.64    & 0.41/1.47   & 9.34/21.16  & 1.37/3.95  \\
{\color[HTML]{680100}InceptionV1*}  & 5.18/11.70   & 0.82/2.63   & 12.73/26.19 & 3.40/8.45  \\ \hline
\end{tabular}
}
\end{table}

\vspace{-17pt}
\begin{table}[h!]
\centering
\resizebox{.47\textwidth}{!}{
\begin{tabular}{ccccc}
\hline
\(M_t\)         & \multicolumn{4}{c}{VGGFce2, InceptionV1}              \\ \hline
Image Priors & \multicolumn{2}{c}{CelebA} & \multicolumn{2}{c}{FFHQ} \\ \hline
Method       & \textit{Base}         & $\textit{Base}_{full}$  & \textit{Base}        & $\textit{Base}_{full}$ \\ \hline
{\color[HTML]{FD6864}InceptionV1*}  & 5.00/11.74   & 0.98/2.96   & 12.07/26.20 & 3.11/7.78  \\
{\color[HTML]{CB0000}InceptionV1*} & 6.36/14.77   & 1.47/4.12   & 15.39/31.01 & 3.36/8.48  \\
{\color[HTML]{680100}InceptionV1*}  & 8.53/18.21   & 3.40/8.21   & 19.14/36.43 & 7.96/16.92 \\ \hline
\end{tabular}
}
\caption{\textbf{Acc@1 / Acc@5 for fine-tuning only the classifier (\textit{Base}) and fine-tuning all parameters ($\textit{Base}_{full}$).} The number of samples is set to {\color[HTML]{FD6864}$2.5K$}, {\color[HTML]{CB0000}$5K$}, and {\color[HTML]{680100}$10K$}. \textbf{*} refers to the surrogate model initialized with a pre-trained face recognition model obtained from the Internet. It can be seen that surrogate models obtained through full parameter fine-tuning suffer a severe drop in accuracy on the private dataset. This indicates that when the sample size is small, fine-tuning all parameters severely degrades the feature extraction capability of the surrogate model.}
\label{table:experiment_full_parameter_2}
\end{table}

\vspace{-16pt}
\subsection{Ablation study on the number of models in the ensemble}
\label{sub:ablation_number}

\vspace{-12pt}
\begin{table}[h!]
\centering
\resizebox{.47\textwidth}{!}{
\begin{tabular}{cclcccc}
\hline
{\color[HTML]{000000} \(M_t\)} &
  \multicolumn{6}{c}{{\color[HTML]{000000} VGGFace2, ResNet50}} \\ \hline
{\color[HTML]{000000} Image Priors} &
  \multicolumn{6}{c}{{\color[HTML]{000000} CelebA}} \\ \hline
{\color[HTML]{000000} Method} &
  \multicolumn{2}{c}{{\color[HTML]{000000} \begin{tabular}[c]{@{}c@{}}\textit{SMILE}\\ $N=3$\end{tabular}}} &
  {\color[HTML]{000000} \begin{tabular}[c]{@{}c@{}}\textit{SMILE}\\ $N=4$\end{tabular}} &
  {\color[HTML]{000000} \begin{tabular}[c]{@{}c@{}}\textit{SMILE}\\ $N=5$\end{tabular}} &
  {\color[HTML]{000000} \begin{tabular}[c]{@{}c@{}}\textit{SMILE}\\ $N=10$\end{tabular}} &
  {\color[HTML]{000000} \begin{tabular}[c]{@{}c@{}}\textit{SMILE}\\ $N=50$\end{tabular}} \\ \hline
{\color[HTML]{FD6864} InceptionV1*} &
  \multicolumn{2}{c}{{\color[HTML]{000000} 8.91/19.40}} &
  {\color[HTML]{9A0000} 9.95/21.29} &
  {\color[HTML]{000000} 9.54/20.96} &
  {\color[HTML]{000000} 9.59/20.81} &
  {\color[HTML]{3531FF} 8.22/19.61} \\
{\color[HTML]{FD6864} EfficientNetB0} &
  \multicolumn{2}{c}{{\color[HTML]{000000} 3.08/7.76}} &
  {\color[HTML]{000000} 3.10/8.09} &
  {\color[HTML]{000000} 3.16/8.09} &
  {\color[HTML]{3531FF} 3.06/7.83} &
  {\color[HTML]{9A0000} 3.21/8.26} \\
{\color[HTML]{CB0000} InceptionV1*} &
  \multicolumn{2}{c}{{\color[HTML]{000000} 9.16/19.67}} &
  {\color[HTML]{3531FF} 8.65/19.43} &
  {\color[HTML]{9A0000} 9.95/21.59} &
  {\color[HTML]{000000} 9.79/20.89} &
  {\color[HTML]{000000} 9.33/19.88} \\
{\color[HTML]{CB0000} EfficientNetB0} &
  \multicolumn{2}{c}{{\color[HTML]{3531FF} 4.75/11.59}} &
  {\color[HTML]{000000} 5.04/12.18} &
  {\color[HTML]{000000} 5.08/12.23} &
  {\color[HTML]{000000} 5.23/12.57} &
  {\color[HTML]{9A0000} 5.32/12.62} \\
{\color[HTML]{680100} InceptionV1*} &
  \multicolumn{2}{c}{{\color[HTML]{000000} 14.27/28.28}} &
  {\color[HTML]{000000} 14.09/27.90} &
  {\color[HTML]{000000} 14.36/28.46} &
  {\color[HTML]{9A0000} 14.93/28.89} &
  {\color[HTML]{3531FF} 12.84/27.09} \\
{\color[HTML]{680100} EfficientNetB0} &
  \multicolumn{2}{c}{{\color[HTML]{000000} 7.13/16.81}} &
  {\color[HTML]{000000} 7.43/17.35} &
  {\color[HTML]{000000} 7.73/17.79} &
  {\color[HTML]{000000} 7.69/17.77} &
  {\color[HTML]{9A0000} 8.75/19.37} \\ \hline
\end{tabular}
}
\label{table:experiment_ablation_number_1}
\end{table}

\vspace{-16pt}
\begin{table}[h!]
\centering
\resizebox{.47\textwidth}{!}{
\begin{tabular}{cccccc}
\hline
{\color[HTML]{000000} \(M_t\)} &
  \multicolumn{5}{c}{{\color[HTML]{000000} VGGFace2, ResNet50}} \\ \hline
{\color[HTML]{000000} Image Priors} &
  \multicolumn{5}{c}{{\color[HTML]{000000} FFHQ}} \\ \hline
{\color[HTML]{000000} Method} &
  {\color[HTML]{000000} \begin{tabular}[c]{@{}c@{}}\textit{SMILE}\\ $N=3$\end{tabular}} &
  {\color[HTML]{000000} \begin{tabular}[c]{@{}c@{}}\textit{SMILE}\\ $N=4$\end{tabular}} &
  {\color[HTML]{000000} \begin{tabular}[c]{@{}c@{}}\textit{SMILE}\\ $N=5$\end{tabular}} &
  {\color[HTML]{000000} \begin{tabular}[c]{@{}c@{}}\textit{SMILE}\\ $N=10$\end{tabular}} &
  {\color[HTML]{000000} \begin{tabular}[c]{@{}c@{}}\textit{SMILE}\\ $N=50$\end{tabular}} \\ \hline
{\color[HTML]{FD6864} InceptionV1*} &
  {\color[HTML]{3531FF} 20.45/38.98} &
  {\color[HTML]{000000} 20.79/39.36} &
  {\color[HTML]{9A0000} 21.08/39.21} &
  {\color[HTML]{000000} 21.02/39.30} &
  {\color[HTML]{000000} 20.79/40.41} \\
{\color[HTML]{FD6864} EfficientNetB0} &
  {\color[HTML]{000000} 7.01/16.55} &
  {\color[HTML]{000000} 7.00/16.73} &
  {\color[HTML]{3531FF} 6.98/16.57} &
  {\color[HTML]{000000} 7.04/16.90} &
  {\color[HTML]{9A0000} 7.18/16.97} \\
{\color[HTML]{CB0000} InceptionV1*} &
  {\color[HTML]{3531FF} 23.78/42.76} &
  {\color[HTML]{9A0000} 25.65/45.57} &
  {\color[HTML]{000000} 23.81/42.57} &
  {\color[HTML]{000000} 24.79/43.82} &
  {\color[HTML]{000000} 25.04/44.10} \\
{\color[HTML]{CB0000} EfficientNetB0} &
  {\color[HTML]{000000} 11.57/25.07} &
  {\color[HTML]{000000} 11.94/25.70} &
  {\color[HTML]{3531FF} 11.41/24.93} &
  {\color[HTML]{9A0000} 12.02/25.76} &
  {\color[HTML]{000000} 11.99/25.86} \\
{\color[HTML]{680100} InceptionV1*} &
  {\color[HTML]{3531FF} 32.29/52.51} &
  {\color[HTML]{000000} 32.47/52.66} &
  {\color[HTML]{9A0000} 34.07/54.39} &
  {\color[HTML]{000000} 33.19/53.70} &
  {\color[HTML]{000000} 33.43/53.77} \\
{\color[HTML]{680100} EfficientNetB0} &
  {\color[HTML]{3531FF} 17.51/34.26} &
  {\color[HTML]{000000} 17.96/35.01} &
  {\color[HTML]{000000} 18.13/35.23} &
  {\color[HTML]{000000} 18.95/36.45} &
  {\color[HTML]{9A0000} 19.80/37.82} \\ \hline
\end{tabular}
}
\label{table:experiment_ablation_number_2}
\end{table}

\vspace{-16pt}
\begin{table}[h!]
\centering
\resizebox{.47\textwidth}{!}{
\begin{tabular}{cclcccc}
\hline
{\color[HTML]{000000} \(M_t\)} &
  \multicolumn{6}{c}{{\color[HTML]{000000} VGGFace2, InceptionV1}} \\ \hline
{\color[HTML]{000000} Image Priors} &
  \multicolumn{6}{c}{{\color[HTML]{000000} CelebA}} \\ \hline
{\color[HTML]{000000} Method} &
  \multicolumn{2}{c}{{\color[HTML]{000000} \begin{tabular}[c]{@{}c@{}}\textit{SMILE}\\ $N=3$\end{tabular}}} &
  {\color[HTML]{000000} \begin{tabular}[c]{@{}c@{}}\textit{SMILE}\\ $N=4$\end{tabular}} &
  {\color[HTML]{000000} \begin{tabular}[c]{@{}c@{}}\textit{SMILE}\\ $N=5$\end{tabular}} &
  {\color[HTML]{000000} \begin{tabular}[c]{@{}c@{}}\textit{SMILE}\\ $N=10$\end{tabular}} &
  {\color[HTML]{000000} \begin{tabular}[c]{@{}c@{}}\textit{SMILE}\\ $N=50$\end{tabular}} \\ \hline
{\color[HTML]{FD6864} InceptionV1*} &
  \multicolumn{2}{c}{{\color[HTML]{000000} 21.84/39.29}} &
  {\color[HTML]{000000} 20.41/37.96} &
  {\color[HTML]{000000} 20.20/36.05} &
  {\color[HTML]{9A0000} 22.11/39.03} &
  {\color[HTML]{3531FF} 20.05/37.80} \\
{\color[HTML]{FD6864} EfficientNetB0} &
  \multicolumn{2}{c}{{\color[HTML]{000000} 9.80/20.74}} &
  {\color[HTML]{3531FF} 8.97/19.11} &
  {\color[HTML]{9A0000} 10.00/21.11} &
  {\color[HTML]{000000} 9.81/21.07} &
  {\color[HTML]{000000} 9.74/20.90} \\
{\color[HTML]{CB0000} InceptionV1*} &
  \multicolumn{2}{c}{{\color[HTML]{3531FF} 20.63/37.44}} &
  {\color[HTML]{000000} 20.88/37.15} &
  {\color[HTML]{000000} 21.32/37.90} &
  {\color[HTML]{9A0000} 21.88/39.66} &
  {\color[HTML]{000000} 21.14/38.82} \\
{\color[HTML]{CB0000} EfficientNetB0} &
  \multicolumn{2}{c}{{\color[HTML]{000000} 15.45/29.81}} &
  {\color[HTML]{000000} 15.36/29.44} &
  {\color[HTML]{000000} 15.53/30.09} &
  {\color[HTML]{3531FF} 15.34/29.24} &
  {\color[HTML]{9A0000} 15.71/30.25} \\
{\color[HTML]{680100} InceptionV1*} &
  \multicolumn{2}{c}{{\color[HTML]{000000} 24.87/42.08}} &
  {\color[HTML]{3531FF} 24.35/41.55} &
  {\color[HTML]{000000} 24.92/42.57} &
  {\color[HTML]{9A0000} 25.13/42.48} &
  {\color[HTML]{000000} 24.70/41.98} \\
{\color[HTML]{680100} EfficientNetB0} &
  \multicolumn{2}{c}{{\color[HTML]{000000} 20.91/37.94}} &
  {\color[HTML]{3531FF} 20.68/37.58} &
  {\color[HTML]{000000} 21.10/37.89} &
  {\color[HTML]{000000} 22.70/39.58} &
  {\color[HTML]{9A0000} 24.06/42.06} \\ \hline
\end{tabular}
}
\label{table:experiment_ablation_number_3}
\end{table}

\begin{table}[h!]
\centering
\resizebox{.47\textwidth}{!}{
\begin{tabular}{cccccc}
\hline
{\color[HTML]{000000} \(M_t\)} &
  \multicolumn{5}{c}{{\color[HTML]{000000} VGGFce2, InceptionV1}} \\ \hline
{\color[HTML]{000000} Image Priors} &
  \multicolumn{5}{c}{{\color[HTML]{000000} FFHQ}} \\ \hline
{\color[HTML]{000000} Method} &
  {\color[HTML]{000000} \begin{tabular}[c]{@{}c@{}}\textit{SMILE}\\ $N=3$\end{tabular}} &
  {\color[HTML]{000000} \begin{tabular}[c]{@{}c@{}}\textit{SMILE}\\ $N=4$\end{tabular}} &
  {\color[HTML]{000000} \begin{tabular}[c]{@{}c@{}}\textit{SMILE}\\ $N=5$\end{tabular}} &
  {\color[HTML]{000000} \begin{tabular}[c]{@{}c@{}}\textit{SMILE}\\ $N=10$\end{tabular}} &
  {\color[HTML]{000000} \begin{tabular}[c]{@{}c@{}}\textit{SMILE}\\ $N=50$\end{tabular}} \\ \hline
{\color[HTML]{FD6864} InceptionV1*} &
  {\color[HTML]{000000} 36.54/58.90} &
  {\color[HTML]{000000} 38.06/58.09} &
  {\color[HTML]{9A0000} 38.50/58.61} &
  {\color[HTML]{000000} 37.05/59.09} &
  {\color[HTML]{3531FF} 34.27/55.75} \\
{\color[HTML]{FD6864} EfficientNetB0} &
  {\color[HTML]{000000} 17.46/33.22} &
  {\color[HTML]{000000} 17.10/33.05} &
  {\color[HTML]{3531FF} 17.07/32.59} &
  {\color[HTML]{000000} 17.97/34.02} &
  {\color[HTML]{9A0000} 18.30/34.36} \\
{\color[HTML]{CB0000} InceptionV1*} &
  {\color[HTML]{000000} 40.17/61.00} &
  {\color[HTML]{3531FF} 37.91/58.81} &
  {\color[HTML]{000000} 39.31/59.47} &
  {\color[HTML]{9A0000} 40.94/61.52} &
  {\color[HTML]{000000} 39.86/60.96} \\
{\color[HTML]{CB0000} EfficientNetB0} &
  {\color[HTML]{3531FF} 24.06/42.86} &
  {\color[HTML]{000000} 24.43/43.13} &
  {\color[HTML]{000000} 24.56/43.54} &
  {\color[HTML]{000000} 25.59/44.71} &
  {\color[HTML]{9A0000} 26.81/45.61} \\
{\color[HTML]{680100} InceptionV1*} &
  {\color[HTML]{000000} 45.12/64.68} &
  {\color[HTML]{000000} 44.78/64.42} &
  {\color[HTML]{000000} 44.59/64.20} &
  {\color[HTML]{9A0000} 46.32/65.82} &
  {\color[HTML]{3531FF} 44.44/64.75} \\
{\color[HTML]{680100} EfficientNetB0} &
  {\color[HTML]{3531FF} 31.54/51.59} &
  {\color[HTML]{000000} 32.61/52.50} &
  {\color[HTML]{000000} 32.43/52.03} &
  {\color[HTML]{000000} 33.19/52/90} &
  {\color[HTML]{9A0000} 34.39/54.77} \\ \hline
\end{tabular}
}
\caption{\textbf{Acc@1 / Acc@5 for surrogate models with $N$ models in the ensemble.} The number of samples is set to {\color[HTML]{FD6864}$2.5K$}, {\color[HTML]{CB0000}$5K$}, and {\color[HTML]{680100}$10K$}. \textbf{*} refers to the surrogate model initialized with a pre-trained face recognition model obtained from the Internet. We highlighted the highest-quality surrogate models under a specific setting in {\color[HTML]{9A0000}red} and the lowest-quality surrogate models in {\color[HTML]{3531FF}blue}. As observed, when the sample size is {\color[HTML]{FD6864}$2.5K$}, setting $N=5$ is more likely to yield higher-quality surrogate models. Additionally, as the sample size increases, the quality of the surrogate models shows a positive correlation with the value of $N$. While using $N=5$ is more likely to produce better surrogate models with 2500 samples, selecting $N=3$ in our main experiments is reasonable. This is because, in the context of black-box MIAs, attackers should not have prior knowledge of the optimal value of $N$. The main experiments demonstrate that even with a suboptimal $N$, \textit{SMILE} can still achieve desirable attack performance. Furthermore, we recommend increasing $N$ as the sample size grows to better account for the greater amount of private information.}
\label{table:experiment_ablation_number_4}
\end{table}

\subsection{Why 2500 queries}
\label{sub:why_2500}
Please refer to \cref{sub:challenge_label_only} and \cref{table:experiment_intersection_1}.

\subsection{Art face as the image prior}
\label{sub:art_face}
Please refer \cref{supp-sec:qualitative_details}.

\subsection{Challenges in the label-only setting}
\label{sub:challenge_label_only}

\begin{table*}[h!]
\centering
\resizebox{.97\textwidth}{!}{

\begin{tabular}{cccc}
\hline
Image priors &
  \multicolumn{2}{c}{CelebA\&FFHQ} &
  Examples \\ \hline
Sampling size &
  Intersection size &
  Proportion &
  The indexes \\ \hline
$40K$ &
  63 &
  21.0\% &
  {[}{\color[HTML]{32CB00}5248}, {\color[HTML]{32CB00}3803}, {\color[HTML]{32CB00}7906}, {\color[HTML]{32CB00}2035}, {\color[HTML]{32CB00}3646}, {\color[HTML]{32CB00}3722}, {\color[HTML]{32CB00}5810}, {\color[HTML]{32CB00}7149}, {\color[HTML]{32CB00}365}, {\color[HTML]{32CB00}5503}, {\color[HTML]{32CB00}273}, {\color[HTML]{32CB00}3795}, {\color[HTML]{32CB00}2086}, {\color[HTML]{32CB00}8488}, {\color[HTML]{32CB00}3772}, {\color[HTML]{32CB00}7800}, {\color[HTML]{32CB00}4551}, {\color[HTML]{32CB00}7148}, {\color[HTML]{32CB00}3791}, {\color[HTML]{32CB00}553}{]} \\
$20K$ &
  61 &
  20.33\% &
  {[}{\color[HTML]{32CB00}5248}, {\color[HTML]{32CB00}3803}, {\color[HTML]{32CB00}2035}, {\color[HTML]{32CB00}7906}, {\color[HTML]{32CB00}3646}, {\color[HTML]{32CB00}3722}, {\color[HTML]{32CB00}5810}, {\color[HTML]{32CB00}7149}, {\color[HTML]{32CB00}365}, {\color[HTML]{32CB00}5503}, {\color[HTML]{32CB00}273}, {\color[HTML]{32CB00}3795}, {\color[HTML]{32CB00}8488}, {\color[HTML]{32CB00}2086}, {\color[HTML]{32CB00}7800}, 3078, 2472, {\color[HTML]{32CB00}3772}, {\color[HTML]{32CB00}7148}, 2309{]} \\
$10K$ &
  60 &
  20.0\% &
  {[}{\color[HTML]{32CB00}5248}, {\color[HTML]{32CB00}3803}, {\color[HTML]{32CB00}7906}, {\color[HTML]{32CB00}3646}, {\color[HTML]{32CB00}3722}, {\color[HTML]{32CB00}2035}, {\color[HTML]{32CB00}5810}, {\color[HTML]{32CB00}7149}, {\color[HTML]{32CB00}5503}, {\color[HTML]{32CB00}273}, {\color[HTML]{32CB00}365}, {\color[HTML]{32CB00}3795}, {\color[HTML]{32CB00}8488}, 2309, {\color[HTML]{32CB00}3772}, {\color[HTML]{32CB00}7800}, {\color[HTML]{32CB00}2086}, {\color[HTML]{32CB00}4551}, {\color[HTML]{32CB00}553}, 2472{]} \\
$5K$ &
  58 &
  19.33\% &
  {[}{\color[HTML]{32CB00}5248}, {\color[HTML]{32CB00}3803}, {\color[HTML]{32CB00}3646}, {\color[HTML]{32CB00}3722}, {\color[HTML]{32CB00}7906}, {\color[HTML]{32CB00}2035}, {\color[HTML]{32CB00}5810}, {\color[HTML]{32CB00}5503}, {\color[HTML]{32CB00}3795}, {\color[HTML]{32CB00}273}, {\color[HTML]{32CB00}7149}, {\color[HTML]{32CB00}3772}, {\color[HTML]{32CB00}8488}, {\color[HTML]{32CB00}4551}, 2309, {\color[HTML]{32CB00}2086}, {\color[HTML]{32CB00}365}, {\color[HTML]{32CB00}7800}, 4506, 3791{]} \\
$2.5K$ &
  47 &
  15.66\% &
  {[}{\color[HTML]{32CB00}5248}, {\color[HTML]{32CB00}3803}, {\color[HTML]{32CB00}3646}, {\color[HTML]{32CB00}3722}, {\color[HTML]{32CB00}2035}, {\color[HTML]{32CB00}7906}, {\color[HTML]{32CB00}5810}, {\color[HTML]{32CB00}5503}, {\color[HTML]{32CB00}273}, {\color[HTML]{32CB00}7149}, {\color[HTML]{32CB00}3772}, {\color[HTML]{32CB00}3795}, {\color[HTML]{32CB00}2086}, {\color[HTML]{32CB00}8488}, {\color[HTML]{32CB00}365}, 1234, {\color[HTML]{32CB00}7800}, 8407, 3078, {\color[HTML]{32CB00}553}{]} \\
$1K$ &
  46 &
  15.33\% &
  {[}{\color[HTML]{32CB00}5248}, {\color[HTML]{32CB00}2035}, {\color[HTML]{32CB00}3722}, {\color[HTML]{32CB00}3646}, {\color[HTML]{32CB00}7906}, {\color[HTML]{32CB00}273}, 3784, 1234, 5758, {\color[HTML]{32CB00}5503}, {\color[HTML]{32CB00}7149}, 8407, {\color[HTML]{32CB00}8488}, 8193, 8227, {\color[HTML]{32CB00}3772}, 4113, 1427, 4896, 5710{]} \\
$0.5K$ &
  32 &
  10.66\% &
  {[}{\color[HTML]{32CB00}5248}, {\color[HTML]{32CB00}2035}, {\color[HTML]{32CB00}3646}, {\color[HTML]{32CB00}5810}, {\color[HTML]{32CB00}3722}, {\color[HTML]{32CB00}8488}, {\color[HTML]{32CB00}7149}, 8193, 8407, 1234, 1427, 7641, {\color[HTML]{32CB00}365}, 7042, 934, 4679, 7886, 741, 4979, 884{]} \\ \hline
\end{tabular}

}
\caption{We set \(M_t\) as ResNet50 pre-trained on VGGFace2. We calculate the intersection of the top 300 dataset-specific attack-sensitive IDs obtained using CelebA and FFHQ as image priors across various sampling sizes, which serve as the general attack-sensitive IDs. It can be observed that a part of IDs are simultaneously easy to be covered under different image priors, and we consider them to be the most attack-sensitive instances in the privacy dataset. Taking the top 20 general attack-sensitive IDs across various sampling sizes as an example, it can be found that the top 20 general vulnerable IDs generally emerge when the sampling size is $2.5K$.}
\label{table:experiment_intersection_2}
\end{table*}

Label-only is a challenging setting for MIAs because the information available to attackers is extremely limited, and the issue is intensified by the large-scale private ID settings. Existing label-only MIAs require at least one sample corresponding to the target ID to be collected before an attack can be launched, serving as an initial point for subsequent optimization processes~\cite{han2023reinforcement} or for training a T-ACGAN~\cite{nguyen2024label}. This is feasible when the number of private IDs is relatively small (e.g., 50/200/530/1000), but for scenarios with a large number of private IDs, even sampling up to $40K$ samples, over 50\% of private IDs still do not receive any samples (as shown in \cref{fig:logits}), meaning attackers cannot obtain any information about these IDs. Existing MIAs cannot be effectively launched, and \textit{SMILE} faces the same issue, as an initial sampling of $2.5K$ covers only a small portion of private IDs, as shown in \cref{table:experiment_cover_labelonly_2}. Therefore, in our setup, launching a label-only MIA for each ID is unfeasible, and we do not wish to increase the number of queries to millions like existing label-only MIAs~\cite{han2023reinforcement,nguyen2024label}. In the label-only setting, we propose a new objective: To compromise as many private IDs as possible with as few queries as necessary. We introduce the concept of \textbf{Attack-sensitive ID}, which includes \textbf{General Attack-sensitive ID} and \textbf{Dataset-specific Attack-sensitive ID}. Attack-sensitive IDs, in the context of label-only MIAs, refer to IDs that receive more samples at initialization, meaning that these IDs are relatively more exposed to attackers. For a specific ID, having access to more samples provides attackers with the opportunity to either directly expose private information or obtain better initial points that are beneficial for subsequent optimization. Dataset-specific attack-sensitive ID refers to attack-sensitive IDs under specific image priors, which are easier to attack under this prior, details in \cref{table:experiment_intersection_1}. General attack-sensitive ID includes IDs that are vulnerable under various image priors and represents the intersection of different dataset-specific attack-sensitive IDs, as \cref{table:experiment_intersection_2}.

We do not perform iterative optimization and only use long-tail surrogate training. We perform white-box MIAs on local surrogate models targeting dataset-specific attack-sensitive IDs, with results presented in ~\cref{table:experiment_labelonly_2}.

\section{Details of experimental setup}
\label{supp-sec:experimental_setup_details}

\paragraph{Datasets \& Models.} Following \cite{an2022mirror}, we target face recognition models that are pre-trained on VGGFace, VGGFace2, and CASIA datasets, all of which are obtained from the Internet. This means that our attack does not need for the private training datasets themselves. However, to demonstrate the quality of the surrogate models (measured as Acc@1 and Acc@5 on the private dataset), extend the pre-trained model architectures used for initializing the surrogate models, and compute KNN Dist and Feat Dist, we need access to the private training datasets. We obtain them from this\footnote{\url{https://www.kaggle.com/datasets/hearfool/vggface2}}\footnote{\url{https://www.kaggle.com/datasets/debarghamitraroy/casia-webface}}, and it is noted that these data are not high-quality original data and have not been strictly aligned with the pre-trained models, leading to lower accuracy of the pre-trained models on the test dataset. We randomly sample 10\% from each dataset as the test dataset to evaluate the pre-trained and surrogate models. Using the remaining 90\% as training data, we pretrain multiple models on various architectures as initializations for surrogate models. It is important to note that in the experiments, the pre-trained models used to initialize the surrogate models and the target pre-trained models are pre-trained on data from different distributions. The details of the models are shown in \cref{table:experiment_model_details}. The pre-trained GAN models used in the experiments to generate high-resolution images are from this\footnote{\url{https://github.com/genforce/genforce}}.

\begin{table}[h!]
\centering
\resizebox{.47\textwidth}{!}{
\begin{tabular}{cccccc}
\hline
Image priors & \multicolumn{5}{c}{CelebA}          \\ \hline
Sampling size &
  {\color[HTML]{CB0000} \(N>10\)} & {\color[HTML]{F56B00} \(10\geq N \geq 5\)} & {\color[HTML]{FFCA2C} \(5>N \geq 2\)} & {\color[HTML]{32CB00} \(N=1\)} & {\color[HTML]{3166FF} \(N=0\)} \\ \hline
$2.5K$         & 0.36\% & 0.94\% & 3.26\%  & 7.31\%  & 88.13\% \\
$5K$           & 1.04\% & 1.74\% & 5.07\%  & 9.73\%  & 82.42\% \\
$10K$          & 2.28\% & 3.07\% & 7.61\%  & 12.16\% & 74.88\% \\
$20K$          & 4.55\% & 4.41\% & 11.19\% & 13.99\% & 65.86\% \\ \hline
\end{tabular}
}
\label{table:experiment_cover_labelonly_1}
\end{table}

\vspace{-16pt}
\begin{table}[h!]
\centering
\resizebox{.47\textwidth}{!}{
\begin{tabular}{cccccc}
\hline
Image priors & \multicolumn{5}{c}{FFHQ}            \\ \hline
Sampling size &
  {\color[HTML]{CB0000} \(N>10\)} & {\color[HTML]{F56B00} \(10\geq N \geq 5\)} & {\color[HTML]{FFCA2C} \(5>N \geq 2\)} & {\color[HTML]{32CB00} \(N=1\)} & {\color[HTML]{3166FF} \(N=0\)} \\ \hline
$2.5K$         & 0.22\% & 0.65\% & 4.36\%  & 10.28\% & 84.49\% \\
$5K$           & 0.61\% & 1.91\% & 7.61\%  & 14.13\% & 75.74\% \\
$10K$          & 1.90\% & 4.08\% & 12.17\% & 15.88\% & 65.97\% \\
$20K$          & 4.67\% & 7.06\% & 16.39\% & 18.05\% & 53.83\% \\ \hline
\end{tabular}
}
\caption{We set \(M_t\) as ResNet50 pre-trained on VGGFace2. $N$ denotes the number of samples in single label. As observed, limited sampling leads to the majority of IDs not receiving samples. This means that attacking each ID is not feasible.}
\label{table:experiment_cover_labelonly_2}
\end{table}

\begin{table*}[h!]
\centering
\resizebox{.97\textwidth}{!}{

\begin{tabular}{cccccccccccc}
\hline
Role &
  Architecture &
  Training dataset &
  Input resolution &
  Classes &
  Source &
  Report Acc@1 &
  Test Acc@1 &
  Epoch &
  Batch size &
  Optimizer &
  Learning rate \\ \hline
\(M_t\)      & VGG16          & VGGFace  & 224*224 & 2622  & ~\cite{VGG16} & 97.22 & -     & -  & -   & -    & -     \\
\(M_t\)      & VGG16BN        & VGGFace  & 224*224 & 2622  & ~\cite{VGG16} & 96.29 & -     & -  & -   & -    & -     \\
\(M_t\)/\(M_s\) &
  ResNet50 &
  VGGFace2 &
  224*224 &
  8631 &
  ~\cite{VGG16} &
  99.88 &
  96.99 &
  - &
  - &
  - &
  - \\
\(M_t\)/\(M_s\) & InceptionV1    & VGGFace2 & 160*160 & 8631  & ~\cite{INCEPTIONv1}                                   & 99.65 & 93.70 & -  & -   & -    & -     \\
\(M_s\)      & InceptionV3    & VGGFace2 & 342*342 & 8631  & -                                   & -     & 95.04 & 20 & 64  & adam & 0.001 \\
\(M_s\)      & MobileNetV2    & VGGFace2 & 224*224 & 8631  & -                                   & -     & 94.47 & 20 & 128 & adam & 0.001 \\
\(M_s\)      & EfficientNetB0 & VGGFace2 & 256*256 & 8631  & -                                   & -     & 96.69 & 20 & 128 & adam & 0.001 \\
\(M_s\)      & Swin-T         & VGGFace2 & 260*260 & 8631  & -                                   & -     & 93.21 & 6  & 20  & adam & 0.001 \\
\(M_t\)/\(M_s\) & InceptionV1    & CASIA    & 160*160 & 10575 & ~\cite{INCEPTIONv1}                                   & 99.05 & 87.31 & -  & -   & -    & -     \\
\(M_t\)      & SphereFace     & CASIA    & 112*96  & 10575 & ~\cite{SPHEREFACE}                                   & 99.22 & -  & -  & -   & -    & -     \\
\(M_s\)      & EfficientNetB0 & CASIA    & 256*256 & 10575 & -                                   & -     & 91.24 & 60 & 128 & adam & 0.001 \\ \hline
\end{tabular}

}
\caption{\textbf{Details of the models.} We are unable to obtain the VGGFace dataset, so we do not test the accuracy, as well as the KNN Dist or Feat Dist in our experiments.}
\label{table:experiment_model_details}
\end{table*}

\begin{table}[h!]
\centering
\resizebox{.47\textwidth}{!}{
\begin{tabular}{ccccc}
\hline
Image priors & \multicolumn{2}{c}{CelebA} & \multicolumn{2}{c}{FFHQ} \\ \hline
Sampling size &
  \begin{tabular}[c]{@{}c@{}}Intersection size\end{tabular} &
  Proportion &
  \begin{tabular}[c]{@{}c@{}}Intersection size\end{tabular} &
  Proportion \\ \hline
$40K$          & 2000       & 100\%         & 2000      & 100\%        \\
$20K$          & 1671       & 83.55\%       & 1717      & 85.85\%      \\
$10K$          & 1526       & 76.3\%        & 1500      & 75.0\%       \\
$5K$           & 1272       & 63.6\%        & 1382      & 69.1\%       \\
{\color[HTML]{FD6864}$2.5K$}         & {\color[HTML]{FD6864}1005}       & {\color[HTML]{FD6864}50.25\%}       & {\color[HTML]{FD6864}1110}      & {\color[HTML]{FD6864}55.5\%}       \\
$1K$           & 736        & 36.8\%        & 814       & 40.7\%       \\
$0.5K$         & 630        & 31.5\%        & 664       & 33.2\%       \\ \hline
\end{tabular}
}
\caption{We set \(M_t\) as ResNet50 pre-trained on VGGFace2. For various sampling sizes, we calculated the top 2000 IDs containing the highest number of samples. We use the top 2000 IDs from a sample size of $40K$ as an approximation of the top 2000 dataset-specific attack-sensitive IDs with the utilized image prior, since it involves substantial sampling. We then analyze the intersection size and proportion of these top 2,000 IDs at reduced sampling sizes with the top 2000 IDs at $40K$ sampling. Notably, when the sampling size is reduced to $2.5K$, the intersection still exceeds 50\%. This indicates that even with a significant reduction in sampling size, $2.5K$ samples can still provide a decent approximation of the image prior. We believe this is sufficient for black-box MIAs, which is why we set the sampling size to $2.5K$. Similarly, when the sampling size is reduced to $0.5K$, the top 2000 IDs still overlap by more than 30\% with those under the $40K$ sampling size. Thus, dataset-specific attack-sensitive IDs begin to manifest even at lower sampling sizes, giving attackers the potential to identify these vulnerable IDs earlier, causing the acceleration of privacy leakage.}
\label{table:experiment_intersection_1}
\end{table}

\paragraph{Hyperparameters of MIAs.}
The settings for the number of queries are shown in \cref{table:motivation_table_2}. For Mirror-w, the optimizer is adam with a learning rate of 0.2; For PPA, the optimizer is adam with a learning rate of 0.005; For RLBMI, we directly use the settings in their open source code. For \textit{SMILE}, in all experiments, the hyperparameters for long-tailed surrogate training are uniformly set to \(\alpha_{ce}=0.15,\alpha_{diversity}=10\), \textit{Top-10 Reweight}, the hyperparameters for gradient-free black-box optimization are set to \(k=1.7\), the optimizer is adam with a learning rate of 0.2. All temperatures \(T\) used for distillation with KL divergence are set to 0.5.

\begin{table}[h!]
\centering
\resizebox{.47\textwidth}{!}{
\begin{tabular}{ccccc}
\hline
\(\mathcal{D}_{priv}\)        & \multicolumn{4}{c}{VGGFace2}                                   \\ \hline
Image Priors   & \multicolumn{4}{c}{CelebA}                                     \\ \hline
\(M_t\)           & \multicolumn{2}{c}{ResNet50} & \multicolumn{2}{c}{InceptionV1} \\ \hline
Method         & Acc@1$\uparrow$         & Acc@5$\uparrow$        & Acc@1$\uparrow$          & Acc@5$\uparrow$          \\ \hline
{\color[HTML]{FD6864}InceptionV1*}   & 22.45         & 44.90        & 16.33          & 38.78          \\
{\color[HTML]{FD6864}EfficientNetB0} & 16.33         & 30.61        & 18.37          & 32.65          \\
{\color[HTML]{CB0000}InceptionV1*}   & 22.45         & 40.82        & 26.53          & 61.22          \\
{\color[HTML]{CB0000}EfficientNetB0} & 24.49         & 48.98        & 26.53          & 61.22          \\
{\color[HTML]{680100}InceptionV1*}   & 16.33         & 30.61        & 28.57          & 59.18          \\
{\color[HTML]{680100}EfficientNetB0} & 20.41         & 34.69        & 36.73          & 57.14          \\
{\color[HTML]{330001}InceptionV1*}   & 24.49         & 40.82        & 32.65          & 65.31          \\
{\color[HTML]{330001}EfficientNetB0} & 24.49         & 55.10        & 38.78          & 69.39          \\ \hline
\end{tabular}
}
\label{table:experiment_labelonly_1}
\end{table}

\begin{table}[h!]
\centering
\resizebox{.47\textwidth}{!}{
\begin{tabular}{ccccc}
\hline
\(\mathcal{D}_{priv}\)        & \multicolumn{4}{c}{VGGFace2}                                   \\ \hline
Image Priors   & \multicolumn{4}{c}{FFHQ}                                       \\ \hline
\(M_t\)           & \multicolumn{2}{c}{ResNet50} & \multicolumn{2}{c}{InceptionV1} \\ \hline
Method         & Acc@1$\uparrow$         & Acc@5$\uparrow$        & Acc@1$\uparrow$          & Acc@5$\uparrow$          \\ \hline
{\color[HTML]{FD6864}InceptionV1*}   & 24.49         & 53.06        & 38.78          & 59.18          \\
{\color[HTML]{FD6864}EfficientNetB0} & 26.53         & 36.73        & 38.78          & 63.27          \\
{\color[HTML]{CB0000}InceptionV1*}   & 28.57         & 55.10        & 34.69          & 69.18          \\
{\color[HTML]{CB0000}EfficientNetB0} & 28.57         & 42.86        & 34.69          & 61.22          \\
{\color[HTML]{680100}InceptionV1*}   & 24.49         & 51.02        & 46.94          & 65.31          \\
{\color[HTML]{680100}EfficientNetB0} & 34.69         & 55.10        & 59.18          & 77.55          \\
{\color[HTML]{330001}InceptionV1*}   & 30.61         & 57.14        & 55.10          & 77.55          \\
{\color[HTML]{330001}EfficientNetB0} & 36.73         & 63.27        & 44.90          & 73.47          \\ \hline
\end{tabular}
}
\caption{The number of samples is set to {\color[HTML]{FD6864}$2.5K$}, {\color[HTML]{CB0000}$5K$}, {\color[HTML]{680100}$10K$}, and {\color[HTML]{330001}$20K$}. \textbf{*} refers to the surrogate model initialized with a pre-trained face recognition model obtained from the Internet. Targeting dataset-specific attack-sensitive IDs, long-tailed surrogate training can effectively obtain private information even in the label-only setting and very limited number of queries.}
\label{table:experiment_labelonly_2}
\end{table}

\paragraph{Evaluation metrics.} 
Following Mirror~\cite{an2022mirror}, we employ two models pre-trained on the same dataset, each serving as the evaluation model for the other, and report the \textbf{Acc@1} and \textbf{Acc@5}; K-Nearest Neighbors Distance (KNN Dist) measures the shortest distance in the feature space between the reconstructed image and the private images of the target ID; Feature Distance (Feat Dist) measures the distance between the feature of the reconstructed image and the average feature of the target ID's private images. The feature distance is the \textit{$l_2$} distance between the outputs from the penultimate layer of the evaluation model. We attack the first 49 IDs of all datasets, and our main experiment in \cref{table:experiment_ffhq_VGGFace_CASIA_2500} is repeated 3 times.

\section{Details of defenses}
\label{supp-sec:defenses_details}
We implement the defenses on MobileNetV2 and Swin Transformer pre-trained on VGGFace2, and the hyperparameters and experimental results are shown in \cref{table:experiment_defense} and \cref{table:experiment_defense_all}. We note that the gradient-free black-box optimization process is severely disturbed when attacking the model under MID defense. We believe that this is caused by the random noise introduced by MID during inference, which makes it difficult for the black-box optimization process to converge. Therefore, for MID, we only use the white-box attack results on the surrogate models.

\begin{table}[h!]
\centering
\resizebox{0.46\textwidth}{!}{
\begin{tabular}{ccccccccccc}
\hline
                       &                                    &       & \multicolumn{2}{c}{PPA} & \multicolumn{2}{c}{Mirror-b} & \multicolumn{2}{c}{RLBMI} & \multicolumn{2}{c}{SMILE} \\ \hline
Defenses               & Hyperparameters                    & Acc   & Acc@1$\uparrow$      & Acc@5$\uparrow$      & Acc@1$\uparrow$         & Acc@5$\uparrow$        & Acc@1$\uparrow$       & Acc@5$\uparrow$       & Acc@1$\uparrow$       & Acc@5$\uparrow$       \\ \hline
                       & {\color[HTML]{9A0000} 0.006, 0.06} & 92.20 (2.27$\downarrow$) & 89.80      & 97.96      & 14.29         & 46.94        & 24.49       & 36.73       & 30.61       & 57.14       \\
\multirow{-2}{*}{BiDO} & {\color[HTML]{00009B} 0.03, 0.3}   & 91.57 (1.63$\downarrow$) & 63.27      & 91.84      & 10.20         & 26.53        & 18.37       & 26.53       & 26.53       & 57.14       \\ \hline
                       & {\color[HTML]{9A0000} 0.005}       & 91.31 (3.16$\downarrow$) & 100.00     & 100.00     & 16.33         & 44.90        & 40.81       & 57.14       & 28.57       & 44.90       \\
\multirow{-2}{*}{MID}  & {\color[HTML]{00009B} 0.005}       & 89.50 (3.70$\downarrow$)  & 91.83      & 97.95      & 6.12          & 14.29        & 26.53       & 42.86       & 6.12        & 12.24       \\ \hline
                       & {\color[HTML]{9A0000} -0.001}      & 92.40 (2.07$\downarrow$) & 83.67      & 93.88      & 14.29         & 38.78        & 20.41       & 32.65       & 20.41       & 38.78       \\
\multirow{-2}{*}{LS}   & {\color[HTML]{00009B} -0.0005}     & 92.48 (0.72$\downarrow$) & 59.18      & 67.35      & 14.29         & 28.57        & 22.45       & 36.73       & 30.61       & 48.98       \\ \hline
                       & {\color[HTML]{9A0000} Block 4}     & 93.82 (0.65$\downarrow$) & 81.63      & 93.88      & 14.29         & 40.82        & 28.57       & 53.06       & 30.61       & 44.90       \\
\multirow{-2}{*}{TL}   & {\color[HTML]{00009B} Block 3}     & 91.98 (1.22$\downarrow$) & 57.14      & 77.55      & 10.20         & 28.57        & 14.29       & 24.49       & 30.61       & 36.73       \\ \hline
\end{tabular}
}
\caption{\textbf{Performance of MIAs under defenses, with FFHQ as the image prior.} The private dataset is VGGFace2, {\color[HTML]{9A0000}red} refers to {\color[HTML]{9A0000}MobileNetV2}, and {\color[HTML]{00009B}blue} refers to {\color[HTML]{00009B}Swin Transformer.}}
\label{table:experiment_defense_all}
\end{table}

\section{The performance of long-tailed surrogate training on the private dataset}
\label{supp-sec:performance_private}
\begin{figure}[h]
  \centering
  \begin{overpic}[width=0.49\textwidth]{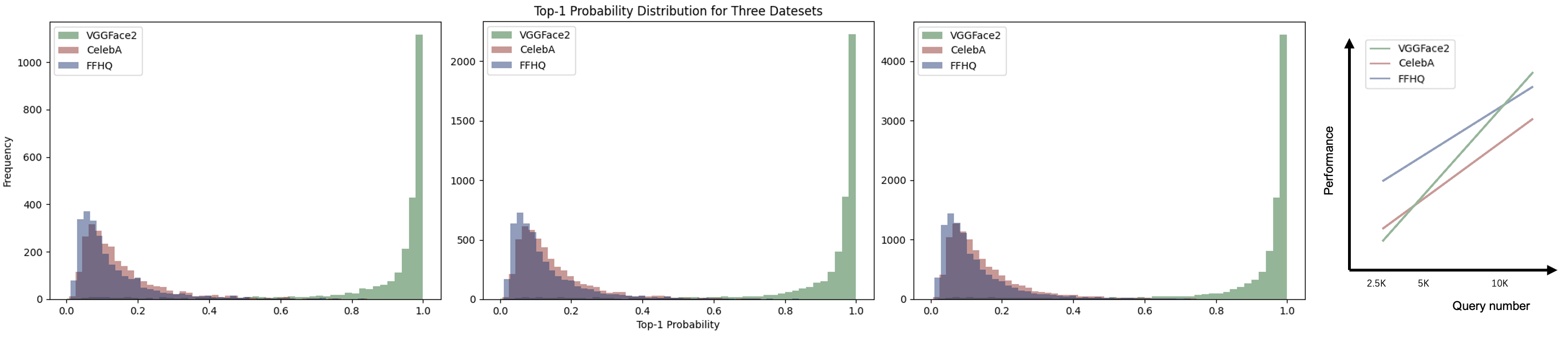}
  
  \put(12.5,17){\footnotesize{{\color[HTML]{FD6864}$2.5K$}}}
  \put(42,17){\footnotesize{{\color[HTML]{CB0000}$5K$}}}
  \put(69,17){\footnotesize{{\color[HTML]{680100}$10K$}}}

  \put(46,-0.2){\footnotesize{(a)}}
  \put(86,-0.1){\footnotesize{(b)}}

  \end{overpic}
  \caption{\textbf{Distribution of Top-1 confidence scores.} The target model is set to VGGFace2,ResNet50.}
  \label{fig:rebuttal_distribution}
\end{figure}

We evaluate surrogate model performance using VGGFace2 as the private dataset with varying sample sizes, as shown in ~\cref{table:R1-Q1}.
When the sample size is set to {\color[HTML]{FD6864}$2.5K$}, surrogate models trained on the {\color[HTML]{036400}private dataset} ($M_s^{priv}$) perform worse than those trained on the public dataset ($M_s^{pub}$). As the sample size increases, $M_s^{priv}$ gradually approach or surpass $M_s^{pub}$. This phenomenon is explained in ~\cref{fig:rebuttal_distribution}(a), 
where the public 
dataset results in lower Top-1 confidence scores, indicating flat model outputs, whereas the private dataset yields higher Top-1 confidence scores.
We believe flat outputs are more beneficial for surrogate models in shaping decision boundaries with limited training data, while higher confidence provides clearer category information when more data is available. The general performance trend of long-tailed surrogate training is shown in ~\cref{fig:rebuttal_distribution}(b).

\begin{table}[h]
\resizebox{0.48\textwidth}{!}{

\begin{tabular}{ccccccccc}
\hline
$D_{priv}$          & \multicolumn{4}{c}{VGGFace2,InceptionV1}              & \multicolumn{4}{c}{VGGFace2,ResNet50}                 \\ \hline
$D$              & CelebA      & FFHQ        & {\color[HTML]{036400}VGGFace2}    & {\color[HTML]{3166FF}VGGFace2}    & CelebA      & FFHQ        & {\color[HTML]{036400}VGGFace2}    & {\color[HTML]{3166FF}VGGFace2}    \\ \hline
{\color[HTML]{FD6864}InceptionV1*}   & 21.84/39.29 & 36.54/58.90 & 11.74/20.36 & 10.83/15.36 & 8.91/19.40  & 20.45/38.98 & 8.23/16.42  & 8.19/16.99  \\ \hline
{\color[HTML]{FD6864}EfficientNetB0} & 9.80/20.74  & 17.46/33.22 & 7.66/13.65  & 5.37/9.36   & 3.08/7.76   & 7.01/16.55  & 3.05/6.82   & 2.79/5.69   \\ \hline
{\color[HTML]{CB0000}InceptionV1*}   & 20.63/37.44 & 40.17/61.00 & 21.49/34.39 & 18.61/26.48 & 9.16/19.67  & 23.78/42.76 & 18.90/27.43 & 17.85/25.95 \\ \hline
{\color[HTML]{CB0000}EfficientNetB0} & 15.45/29.81 & 24.06/42.86 & 13.86/22.55 & 9.18/15.43  & 4.75/11.59  & 11.57/25.07 & 5.09/9.84   & 4.98/9.54   \\ \hline
{\color[HTML]{680100}InceptionV1*}   & 24.87/42.08 & 45.12/64.68 & 39.64/56.82 & 25.84/38.18 & 14.27/28.28 & 32.29/52.51 & 33.24/45.67 & 27.98/40.03 \\ \hline
{\color[HTML]{680100}EfficientNetB0} & 20.91/37.94 & 31.54/51.59 & 27.14/41.88 & 20.04/30.92 & 7.13/16.81  & 17.51/34.26 & 12.02/20.36 & 10.26/18.15 \\ \hline
\end{tabular}

}
\caption{We set the sample sizes to {\color[HTML]{FD6864}$2.5K$}, {\color[HTML]{CB0000}$5K$}, and {\color[HTML]{680100}$10K$}. {\color[HTML]{036400}VGGFace2} refers to training surrogate models on outputs of the target model using private dataset, while {\color[HTML]{3166FF}VGGFace2} refers to using the hard labels corresponding to the private dataset.}
\label{table:R1-Q1}
\end{table}

\section{More results about Long-tailed Learning}
\label{supp-sec:more_results}
\begin{figure}[h]
  \centering
  \begin{overpic}[width=0.29\textwidth]{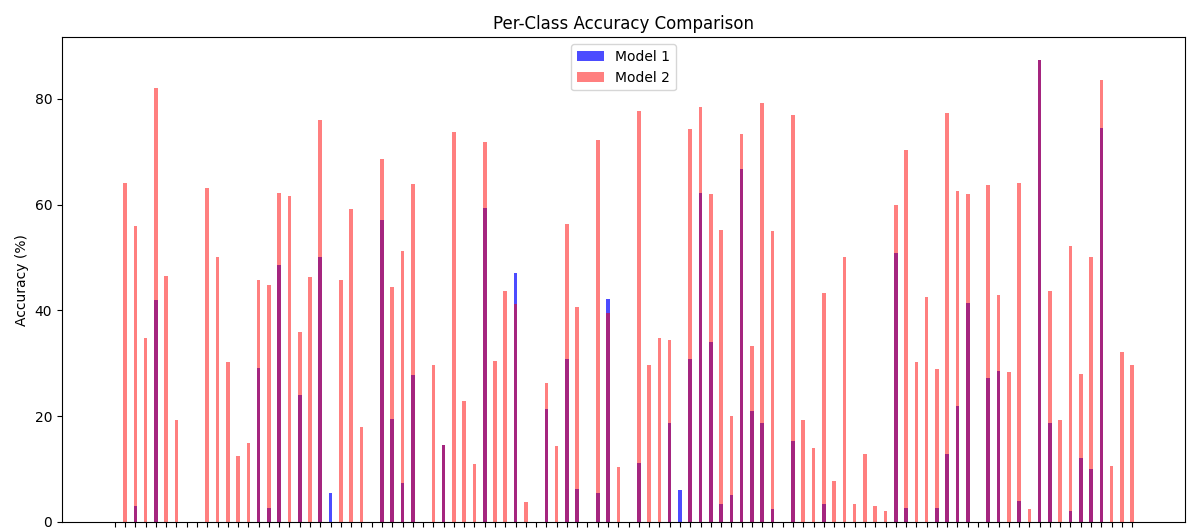}

  \put(5.1,38){\tiny{Target:VGGFace2,InceptionV1}}
  \put(5.3,35){\tiny{Pretrain:CAISA,InceptionV1}}
  \put(5.3,32){\tiny{Auxiliary:FFHQ}}
  \put(5.3,29){\tiny{Query Number:2.5K}}


  \end{overpic}
  \caption{Model 1 is from \textit{Base}, Model 2 is from long-tailed surrogate training. The figure shows the model's performance on the first 100 IDs.}
  \label{fig:R1-Q3}
\end{figure}

A possible strategy to alleviate the long-tail issue is to use auxiliary priors with better diversity (as discussed in Section 3.2.). From the attacker's perspective, our goal is not to rely on better priors or larger sample sizes but to extract more information from extreme long-tail distributions to improve surrogate models. This alleviates the performance degradation of surrogate models caused by the long-tail issue, but does not resolve the long-tail issue itself. We further show the boost that long-tailed surrogate training brings to classification, shown in ~\cref{fig:R1-Q3}. The overall performance improvement can be clearly observed.

\section{More experiments under CASIA pre-trained models}
\label{supp-sec:CASIA_pretrained}
In MIAs, IID refers to splitting a dataset into private and auxiliary parts (both highly aligned). 
We chose VGGFace2 pre-trained models for their diverse architectures, which help validate our method’s robustness. The data distributions of VGGFace and VGGFace2 are relatively close, which may cause concern. Therefore,  we add experiments under CASIA pre-trained models (\cref{table:R3-Q2}). We believe that the similar attack performance is due to the alignment and distribution differences between the target model/pre-trained model and the synthetic data, which weakens the impact of the pre-trained model's training data.

\begin{table}[h]
\centering
\resizebox{0.48\textwidth}{!}{

\begin{tabular}{ccccccccc}
\hline
$D_{priv}$                 & \multicolumn{8}{c}{VGGFace}                                                                                                          \\ \hline
$D_{pub}$                  & \multicolumn{4}{c}{CelebA}                                        & \multicolumn{4}{c}{FFHQ}                                         \\ \hline
$M_t$                    & \multicolumn{2}{c}{VGG16}       & \multicolumn{2}{c}{VGG16BN}     & \multicolumn{2}{c}{VGG16}      & \multicolumn{2}{c}{VGG16BN}     \\ \hline
Method                  & Acc@1          & Acc@5          & Acc@1          & Acc@5          & Acc@1          & Acc@5         & Acc@1          & Acc@5          \\ \hline
\textbf{Average}        & 71.28          & 85.91          & 68.56          & 79.79          & 77.89          & 90.02         & 66.3           & 84.23          \\ \hline
\textbf{EfficientNetB0} & 69.39 ±   3.33 & 82.31 ±   1.92 & 70.75 ±   0.96 & 76.87 ±   0.96 & 80.95 ±   3.46 & 93.20 ±   0.96 & 68.03 ±   2.54 & 87.07 ±   1.92  \\ \hline
\textbf{InceptionV1*}   & 65.99 ±   2.54 & 83.67 ±   1.66 & 64.62 ±   2.54  & 82.31 ±   2.54 & 78.91 ±   3.46  & 91.16 ±   0.96 & 66.66 ±   3.85  & 80.95 ±   2.54 \\ \hline
\end{tabular}

}
\caption{\textbf{Results under CASIA pre-trained models.} \textbf{Average} is the mean result across different architectures of VGGFace2 pre-trained models.}
\label{table:R3-Q2}
\end{table}

\newpage
\section{More qualitative results}
\label{supp-sec:qualitative_details}

\begin{figure}[h!]
\hspace{1.5cm}
\begin{overpic}[width=0.9\textwidth]{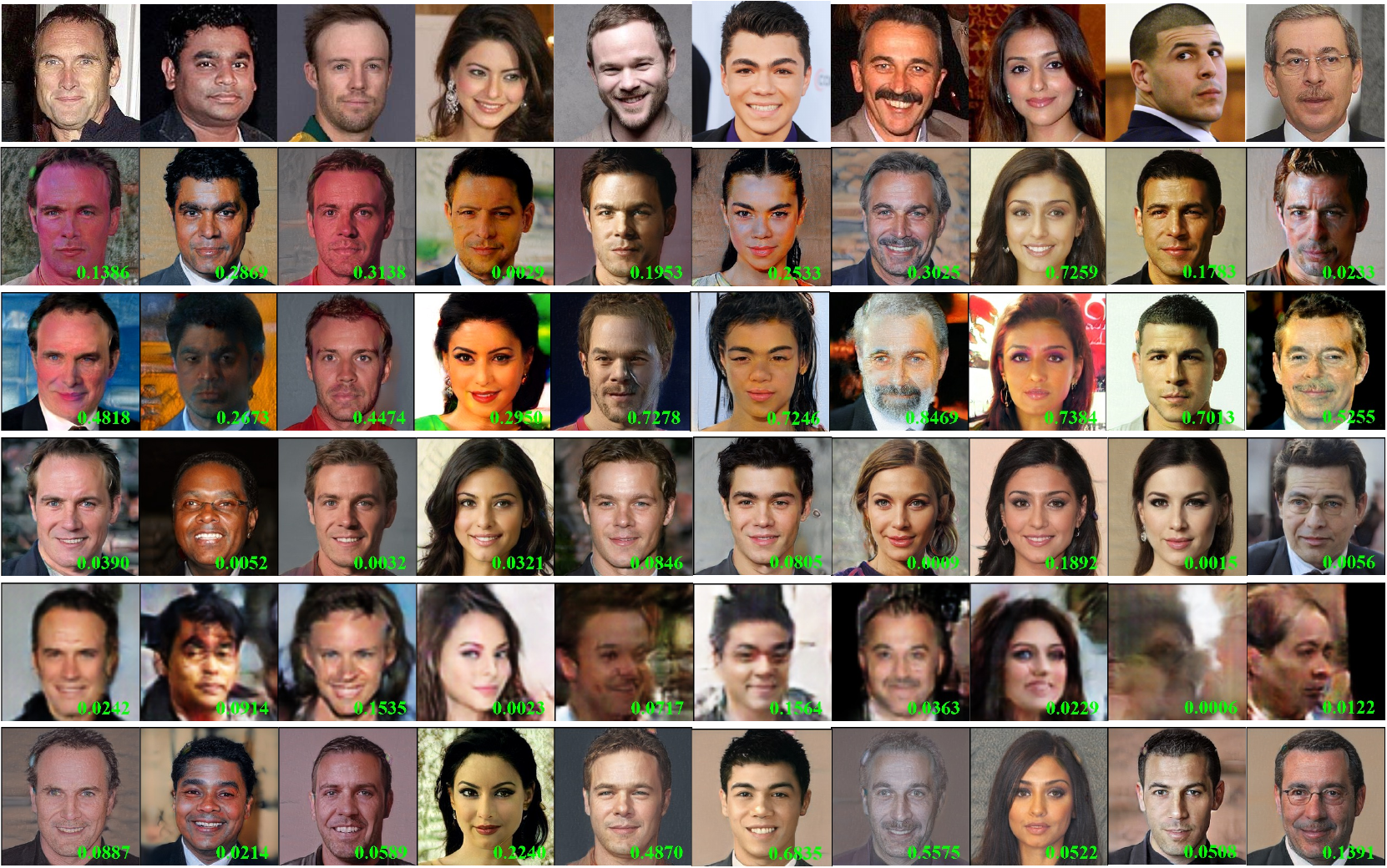}

  \put (-8.0,58) {Ground}
  \put (-7.0,55) {Truth}

  \put (-9.0,46) {Mirror-w}
  \put (-7.0,36) {PPA}

  \put (-9.0,26) {Mirror-b}
  \put (-8.0,15) {RLBMI}
  \put (-8.0,5) {SMILE}
  
  \put (35.0,-3){VGGFace2 / ResNet50 / CelebA}
  \put (38.5, -6){(\(D_{priv}\) / \(M_t\) / Image prior)}

  \end{overpic}
  \label{fig:qualitative_results_1}
\end{figure}


\begin{figure}[h!]
\hspace{1.5cm}
\begin{overpic}[width=0.9\textwidth]{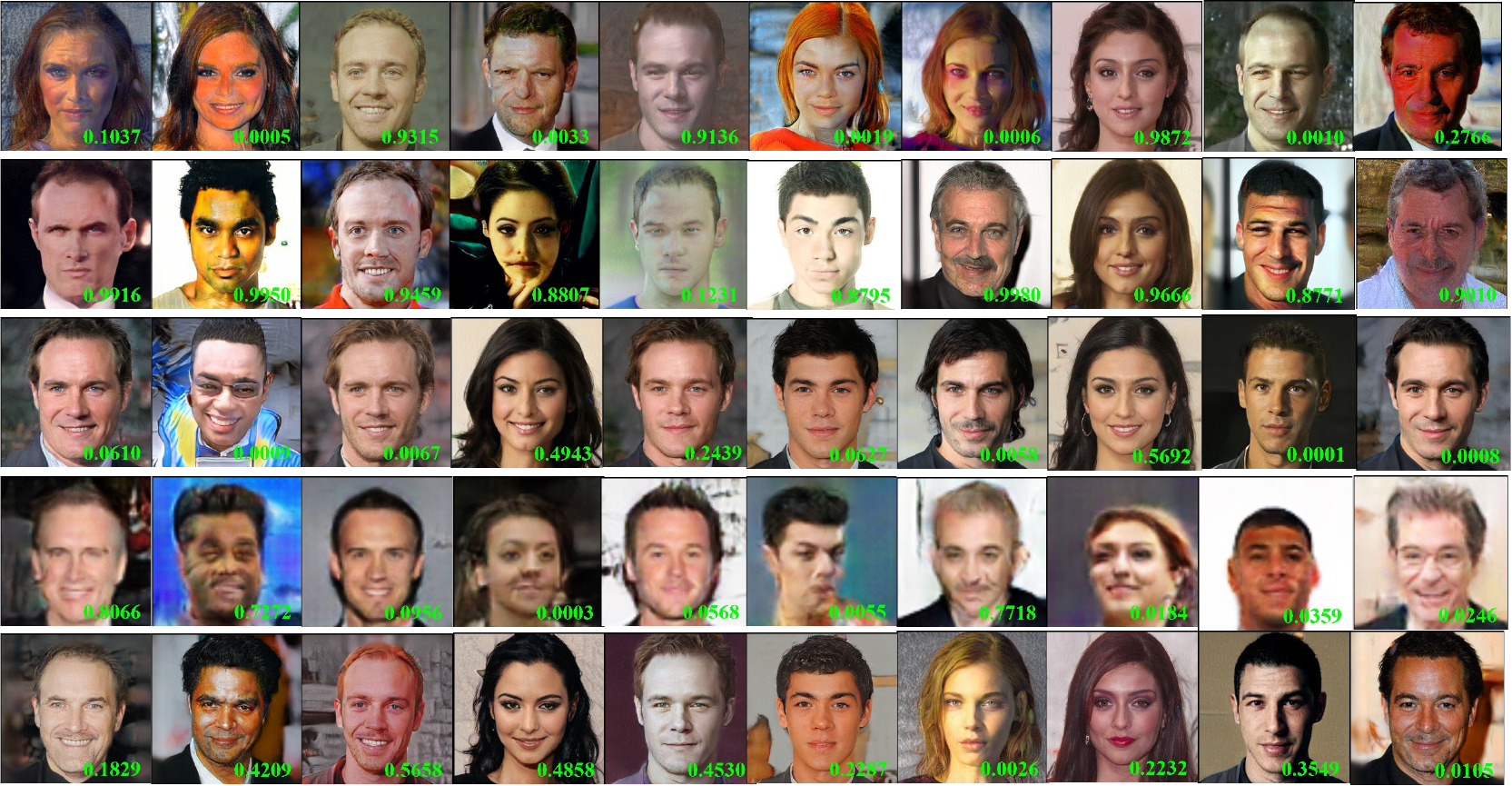}


  \put (-9.0,46) {Mirror-w}
  \put (-7.0,36) {PPA}

  \put (-9.0,26) {Mirror-b}
  \put (-8.0,15) {RLBMI}
  \put (-8.0,5) {SMILE}

  \put (34.0,-3){VGGFace2 / InceptionV1 / CelebA}

  \end{overpic}
  \label{fig:qualitative_results_2}
\end{figure}

\clearpage

\begin{figure}[h!]
\hspace{1.5cm}
\begin{overpic}[width=0.9\textwidth]{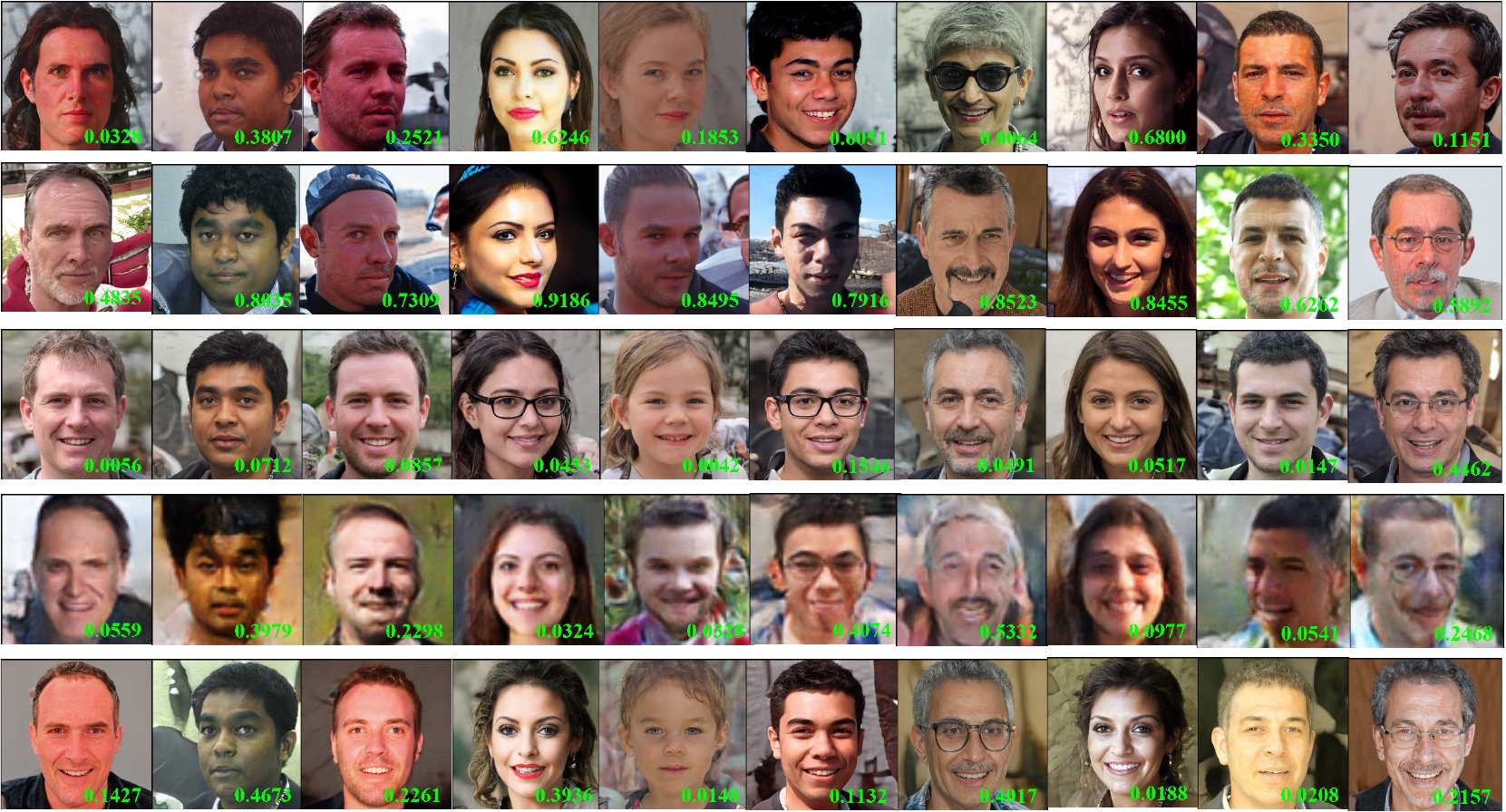}

  \put (-9.0,49) {Mirror-w}
  \put (-7.0,38) {PPA}

  \put (-9.0,26) {Mirror-b}
  \put (-8.0,15) {RLBMI}
  \put (-8.0,5) {SMILE}

  \put (35.5,-3){VGGFace2 / ResNet50 / FFHQ}

  \end{overpic}
  \label{fig:qualitative_results_3}
\end{figure}


\begin{figure}[h!]
\hspace{1.5cm}
\begin{overpic}[width=0.9\textwidth]{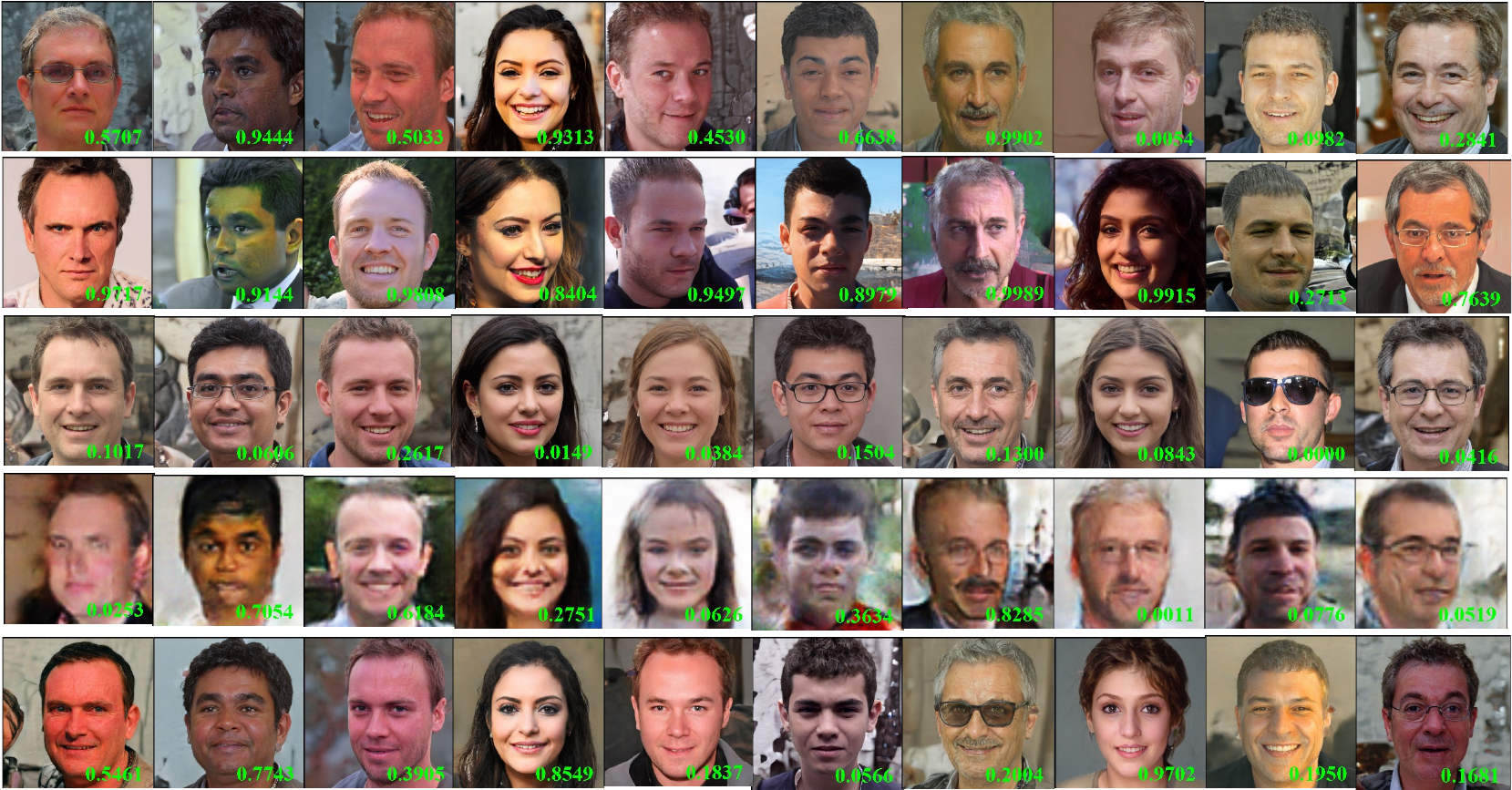}

  \put (-9.0,46) {Mirror-w}
  \put (-7.0,36) {PPA}

  \put (-9.0,26) {Mirror-b}
  \put (-8.0,15) {RLBMI}
  \put (-8.0,5) {SMILE}

  \put (34,-3){VGGFace2 / InceptionV1 / FFHQ}

  \end{overpic}
  \label{fig:qualitative_results_4}
\end{figure}

\clearpage










\begin{figure*}[h!]
\centering
\begin{overpic}[width=0.9\textwidth]{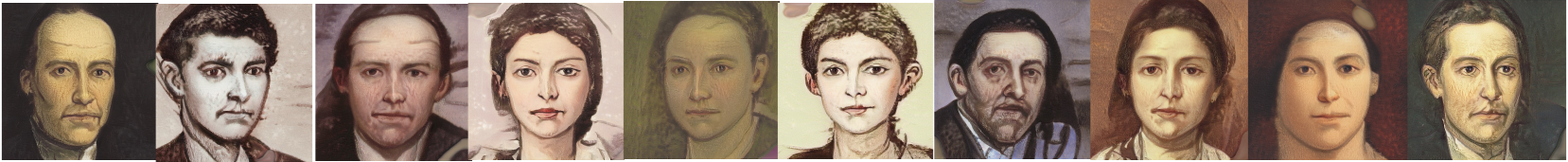}

  \put (-8.0,5) {SMILE}

  \put (35,-3){VGGFace2 / ResNet50 / Artface}

  \end{overpic}

\vspace{22pt}
\begin{overpic}[width=0.9\textwidth]{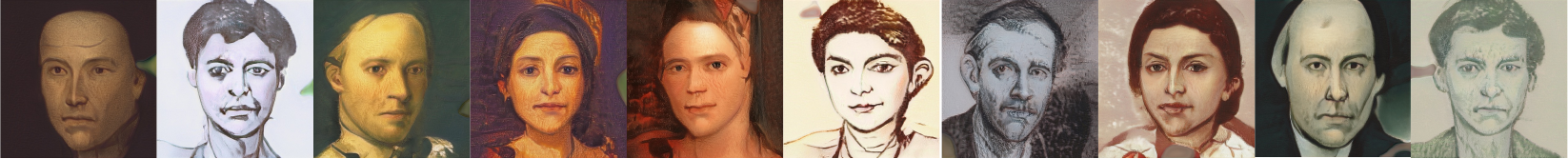}

  \put (-8.0,5) {SMILE}
  \put (34.2,-3){VGGFace2 / InceptionV1 / Artface}

  \end{overpic}
  \vspace{10pt}
  \caption{\textbf{More qualitative results.} The surrogate model used by \textit{SMILE} is InceptionV1 pre-trained on CASIA.}
  \label{fig:qualitative_results_6}

\end{figure*}

\end{document}